\pdfoutput=1

\documentclass[a4paper,11pt]{article}

\usepackage{jheppub}

\usepackage{graphicx} % Required for inserting images
\usepackage{amsmath, amssymb, amsfonts, mathtools}
\usepackage{amsthm}
\usepackage{slashed}
\usepackage{hyperref}
\usepackage{multirow}
\usepackage{xspace}
\usepackage{tikz}
\usetikzlibrary{decorations.markings, decorations.pathreplacing, arrows.meta}
\tikzset{branchcut/.style={thick, red, densely dashed}}

\newcommand{\nn}{\nonumber}
\newcommand{\df}{\mathrm{d}}

\newcommand{\ra}{\rightarrow}

\newcommand{\bb}[1]{\mathbb{#1}}

\newcommand{\Ord}[1]{\mathcal{O} \! \left(#1\right)}

\newcommand{\as}{\alpha_{\rm s}}

\newcommand{\ga}{\gamma}
\newcommand{\de}{\delta}

\newcommand{\eps}{\epsilon}
\newcommand{\la}{\lambda}

\newcommand{\cE}{\mathcal{E}}
\newcommand{\cM}{\mathcal{M}}

\newcommand{\cN}{\mathcal{N}}

\newcommand{\cK}{\mathcal{K}}
\newcommand{\cO}{\mathcal{O}}

\newcommand{\cD}{\mathcal{D}}

\newcommand{\vev}[1]{\langle #1 \rangle}
\newcommand{\bra}[1]{\langle #1 \rvert}
\newcommand{\ket}[1]{\lvert #1 \rangle}

\newcommand{\ee}{{e^+e^-}}

\newcommand{\Tr}{{\rm Tr }\, }
\newcommand{\Pexp}{{\rm Pexp }\, }

\newcommand{\nh}{\hat{n}}

\newcommand{\DLz}{\Delta_L^0}
\newcommand{\PSL}{\widetilde{{\rm PS}}}
\newcommand{\PS}{{\rm PS}}
\newcommand{\EG}{{\rm EG}}
\newcommand{\EGL}{\widetilde{{\rm EG}}}

\newcommand{\id}{\mathbb{I}}
\DeclareMathOperator*{\sumint}{\textstyle \sum\!\!\!\!\!\!\! \displaystyle\int}

\newcommand{\cH}{\mathcal{H}}

\allowdisplaybreaks[2]

\title{What Do Event Generators Compute?\\ Flux Functionals from Detector Operators}

\author[a]{Jack Holguin,}
\affiliation[a]{Department of Physics and Astronomy,
University of Manchester, \\ Manchester M13 9PL, United Kingdom}
\emailAdd{jack.holguin@manchester.ac.uk}

\abstract{
Every physical prediction of an event generator is a statement about the hadronic fluxes that can be recorded by particle detectors. By contrast, the formal backbone of modern event-generator precision is provided by the unobservable partonic final states of the parton shower, with hadronisation included as a model. We construct the probability distribution of the hadronic fluxes non-perturbatively, directly from detector operators acting on asymptotic hadronic states, and encode it in an event generator source functional. This distribution contains every measurement a detector can make, and is therefore what a generator must compute with precision, including the effects of hadronisation. Restricting the functional to the energy flow yields the parton shower source functional, which is infrared and collinear safe and contains every calorimetric observable, making it a natural object against which to assess parton-shower accuracy. We compute its logarithmic structure at one loop in $e^+e^-$ annihilation, finding it directly sensitive to kinematic recoil and hence to the recoil scheme of a parton shower. The parton shower functional depends on a continuum of low scales, one for each resolved direction, and in a toy model its logarithms cannot be entirely resummed within a factorisation whose evolution derives only from the high-scale physics. Our operator-based formulation opens the door to the study of event-generator-like systems in theories beyond QCD, including conformal theories such as $\mathcal{N}=4$ super-Yang--Mills theory, and thereby offers a new route to non-perturbative insight into event generation.
}

\begin{document}
\maketitle

\section{Introduction}

In the past couple of decades, event generation has undergone several dramatic improvements, leading to a general notion of systematically higher accuracy event generation. An important early milestone was the widespread implementation of next-to-leading-order (NLO) matrix elements for hard-scattering processes, first through the MC@NLO method \cite{Frixione:2002ik} and shortly followed by the POWHEG method \cite{Nason:2004rx}. By the mid-2010s, NLO event generators had become the standard for most common collider processes \cite{Buckley:2011ms,Platzer:2011bc,Bierlich:2022pfr,Sherpa:2024mfk,Bellm:2025pcw}. More recently, significant advances have been made in the understanding of parton showers \cite{Dasgupta:2018nvj,Dasgupta:2020fwr,Forshaw:2020wrq,Nagy:2020rmk,Herren:2022jej,vanBeekveld:2024wws}, which dress fixed-order hard processes with arbitrarily large multiplicities of QCD radiation. Together, these developments have placed event generation firmly on the path away from phenomenological modelling and towards systematic, well-controlled QCD simulations of generic collider processes. In this context, the hadronisation stage of event generation remains the final frontier \cite{Hoang:2018zrp,Hoang:2024nqi,Holguin:2026trd}.

Progress in parton showers has been driven along multiple frontiers. By simulating the large multiplicities of radiation that dress an event, a parton shower provides, for a given observable, a Monte Carlo evaluation of the resummation of the large logarithms that appear in the corresponding cross section. Consequently, while it has historically been challenging to define the general-purpose accuracy of a parton shower, one can unambiguously define the accuracy it achieves for a particular observable in terms of the logarithmic accuracy of the associated resummation (leading logarithmic, next-to-leading logarithmic, and so on). This insight has been exploited with great success by the PanScales collaboration \cite{Dasgupta:2020fwr,Hamilton:2020rcu,vanBeekveld:2022zhl}. The PanScales approach has been to consider a set of theoretically calculable observables, each sensitive to different properties of the parton shower, and to verify numerically that the proposed shower algorithms achieve a given logarithmic accuracy across the set. Provided that the selection of observables is sufficiently broad, one is then justified in trusting that the parton shower attains the corresponding level of general-purpose logarithmic accuracy. Combined with modern developments in the implementation of NLO hard-scattering processes \cite{Hamilton:2023dwb,vanBeekveld:2025lpz}, this programme has recently culminated in the demonstration of next-to-next-to-leading logarithmic (NNLL) accuracy across a reasonably broad class of $e^+e^-$ observables \cite{vanBeekveld:2024wws}, while NLL accuracy is gradually becoming the new standard \cite{Dasgupta:2020fwr,Forshaw:2020wrq,Nagy:2020rmk,Herren:2022jej,Preuss:2024vyu,Holguin:2026trd}.

Concurrently, an orthogonal approach to understanding parton showers has also attracted attention from a number of smaller collaborations. This approach has focused on establishing the formal foundations of parton showering through the study of the infrared limits of scattering amplitudes \cite{Angeles-Martinez:2015rna,Nagy:2017ggp,Forshaw:2019ver,Nagy:2019pjp,Forshaw:2020wrq,DeAngelis:2020rvq,Nagy:2020rmk,Platzer:2022jny}. While these collaborations initially pursued different approaches, they have gradually converged on a consistent picture in which the role of a parton shower is to realise the renormalisation-group evolution of a hard-scattering process from a high scale, where it can be computed using fixed-order perturbation theory, to a low scale, where hadronisation takes over \cite{Nagy:2017ggp,Nagy:2020rmk,Platzer:2022jny,Becher:2023mtx}. Where approaches still differ is in how this evolution is formulated. The renormalisation-group evolution of the amplitudes is highly non-trivial because the amplitudes are \emph{completely exclusive} and therefore the evolution is off-diagonal in final-state multiplicity.

The goal of this paper is to explore an approach to the formulation of event generation that lies, in spirit, between the two aforementioned approaches. The motivation is to address several outstanding questions that arise within each framework: \\
Regarding the ``observables approach'' \cite{Dasgupta:2020fwr,Hamilton:2020rcu,vanBeekveld:2022zhl,Forshaw:2020wrq,Nagy:2020rmk,Herren:2022jej,Preuss:2024vyu,Holguin:2026trd}, 
\begin{enumerate} 
    \item How large a set of observables is required to establish that a given level of accuracy has been achieved in a genuinely general-purpose sense? 
    \item How can this approach be extended to processes or observables for which precise resummation has not yet been established independently, such as more complicated exclusive processes or intrinsically non-perturbative observables? 
\end{enumerate} 
Regarding the ``amplitudes approach'' \cite{Angeles-Martinez:2015rna,Nagy:2017ggp,Forshaw:2019ver,Nagy:2019pjp,Forshaw:2020wrq,DeAngelis:2020rvq,Nagy:2020rmk,Platzer:2022jny}, 
\begin{enumerate} 
    \item When working directly with exclusive amplitudes, $\as$ is the only expansion parameter available for a systematic perturbative treatment; logarithms and singularities are not yet fully exposed since the real-emission phase space remains differential. Consequently, it is non-trivial to relate the resulting accuracy to that achieved for a given observable, where both logarithmic and fixed-order accuracies are relevant. This issue already appears at relatively modest orders. For example, it is known that NLL accuracy, even for simple processes, requires the reconstruction of the CMW coupling \cite{Catani:1990rr,Dasgupta:2020fwr}, thereby incorporating some $\mathcal{O}(\as^2)$ effects within a shower whose remaining ingredients are otherwise only accurate to $\mathcal{O}(\as)$. 
    \item QCD amplitudes are mostly studied as fundamentally perturbative objects. Consequently, the evolution is only constrained by perturbation theory. 
\end{enumerate}
At its core, the tension between these two approaches concerns the information they are capable of encoding.\footnote{It is not our intention to be reductive. Whilst a systematic observables approach was arguably pioneered by the PanScales collaboration, PanScales has also made extensive use of matrix-element tests \cite{Dasgupta:2018nvj,Hamilton:2020rcu,Hamilton:2021dyz,Karlberg:2021kwr,vanBeekveld:2022zhl}. Likewise, collaborations studying amplitude evolution have presented results demonstrating the reproduction of particular observables \cite{Forshaw:2020wrq,DeAngelis:2020rvq,Nagy:2020rmk,Nagy:2020dvz,Forshaw:2025fif,Forshaw:2025bmo}.} By focusing on only a finite set of representative observables, the ``observables approach'' necessarily provides incomplete information and therefore does not fully constrain the dynamics of an event generator. In contrast, the ``amplitudes approach'' seeks to constrain aspects of partonic dynamics that are not directly experimentally accessible by working with the unphysical $n$-parton phase space. 

The goal of this manuscript is to present a new formulation of what event generators should compute: one that is non-perturbative from the outset and therefore offers a route towards studying hadronisation as well as parton-shower accuracy. The approach we take is to define an event generator as an algorithmic solution for computing the physical distribution of fluxes measured by a detector.\footnote{This definition is idealised. We do not account for detector acceptance or detector response. Likewise, we work within pure QCD and treat the detectors as being infinitely far from the collision point, so that all measured hadrons are stable.} In this sense, the ``observables approach'' can be viewed as a limiting case of the ``flux approach'' introduced here, while the unphysical information retained in the ``amplitudes approach'' is explicitly integrated out, thereby directly facilitating, for example, KLN cancellations.

In this manuscript, we introduce two key quantities: the event generator source functional ($\EGL$) and the parton shower source functional ($\PSL$). $\EGL$ is the probability distribution for observing, in a collider detector, a given flux carrying a specified set of hadronic quantum numbers and momentum fluxes. $\PSL$ is obtained by marginalising $\EGL$ over all fluxes that are not infrared- and collinear-safe (IRC safe). Consequently, it is $\EGL$ that an event generator \emph{must} be able to compute with precision, since it is this distribution that is measured experimentally. Any difference between two event-generator constructions that does not modify $\EGL$ must therefore be regarded as a scheme choice. In contrast, $\PSL$ defines the class of measurements that can, up to power corrections, be computed by treating final-state radiation perturbatively. As such, it is a natural object with respect to which questions of generalised logarithmic accuracy may be posed. Further marginalisation of $\PSL$ recovers the ``observables approach'' for sets of IRC-safe observables. 

In this work, we present the formulation of $\EGL$ and $\PSL$ in terms of hadronic detector operators. We highlight key properties of both functionals and demonstrate that this formulation admits a natural probabilistic interpretation, ideal for event generation. We additionally compute the logarithmic dependence of $\PSL$ perturbatively at one loop, highlighting its properties and direct sensitivity to kinematic recoil. We leave to future work a further study of the all-orders structure of $\EGL$ and $\PSL$, as well as the realisation of the resummation of the logarithms they contain through the application of a parton-shower algorithm.

This paper is laid out as follows. In section~\ref{sec:measurements} we construct detector operators from asymptotic hadronic modes and identify the commuting class that admits joint readings. From these we define $\PSL$, establish its infrared and collinear safety, positivity and analyticity, and show how any calorimetric observable is obtained from it. We then generalise to $\EGL$. In section~\ref{sec:oneloop} we compute the logarithmic terms of $\PSL$ at one loop for a general twice-differentiable source and give an example with the thrust distribution. In section~\ref{sec:toy} we study the all-orders structure of $\PSL$ using a toy-model of factorisation, relate it to the factorisation of non-global observables, and argue that renormalisation-group evolution of the hard functions alone cannot resum all of its large logarithms. We conclude in section~\ref{sec:conclusions}.

\section{Measurements of asymptotic fluxes}
\label{sec:measurements}

This section develops the central objects upon which the remainder of this work is built. To construct these objects, we must first discuss carefully the measurement of asymptotic fluxes. We do so using the detector-operator construction \cite{Sveshnikov:1995vi,Korchemsky:1997sy,Korchemsky:1999kt,Belitsky:2001ij,Hofman:2008ar,Caron-Huot:2022eqs,Chang:2025zib,Chang:2025kgq,Gonzalez:2026aha}, which we first review and subsequently extend. Using these detector operators, we define the event generator ($\EGL$) and parton shower ($\PSL$) source functionals. Both are defined explicitly in terms of hadronic matrix elements of detector operators. We begin with the parton shower functional, which is the simpler of the two and describes only the energy flux. It is infrared and collinear safe and is therefore the object that admits a purely perturbative leading-power description, the description that a parton shower is intended to provide. We then generalise this construction to $\EGL$, which further resolves the flux into hadron species and spin states.

To simplify matters, we will work throughout with $\ee\ra\ga^*\ra\text{hadrons}$ at $q^2=Q^2$, in the
centre-of-mass frame, so that the final state is created by the electromagnetic current,
\begin{align}
	\ket{\psi}\;=\;\eps_\mu(q)\int\df^4x\;e^{iq\cdot x}\,J^\mu(x)\,\ket{0}\,,
	\qquad \vev{\psi|\psi}=\sigma_{\rm tot}(Q)\,,
	\label{eq:state}
\end{align}
the second equation fixing the normalisation. Additionally, for our purposes, we will define a collider detector as a detector that simultaneously reports the readings in each of its detector cells.

\subsection{Detectors}
\label{sec:detectors}

We will now start by formulating the operators that lie at the core of our construction.

The operators we require are those whose eigenvalues correspond to the measurements we perform with a collider detector. We imagine that these measurements take place at future asymptotic infinity, long after the interaction has ceased. Consequently, the relevant operators are constructed from asymptotic \emph{out} modes describing stable particles. These modes therefore correspond to hadronic degrees of freedom, and the fields describing them are defined within an effective field theory of the hadronic degrees of freedom. Placing the measurement at future asymptotic infinity provides three important simplifications. (i) The algebra of the out modes is that of a free field, and is therefore exact even in the interacting theory, allowing detector operators to be manipulated without approximation. (ii) Operators constructed from these modes are diagonal on out states. (iii) All interacting dynamics resides in the amplitude for producing a given out state, which is the only point at which the state in equation~\eqref{eq:state} enters. As a consequence of these simplifications, we do not need to understand the dynamics of the hadronic effective field theory in order to define the detector operators. We begin with the simplest example: an idealised calorimeter that measures the energy flux passing through it. Then we will generalise.

An out mode is labelled by the species and spin state of the hadron it carries, as well as by the energy and direction with which that hadron arrives. We label the out-state kinematics as follows. Points on the celestial sphere are identified by a unit three-vector $\nh$, and $n^\mu(\nh)=(1,\nh)$ denotes the null direction pointing towards $\nh$. We define the angular delta function to carry two arguments, $\de^{(2)}(\nh,\nh')$, since $\hat n - \hat n'$ is not a point on the sphere. A hadron of species $h$ and mass $m_h$, arriving from the direction $\nh$ with energy $\omega$, carries momentum 
$$ p^\mu=\big(\omega,|\vec p\,|\,\nh\big), $$ 
with 
$$ |\vec p\,|=\sqrt{\omega^{2}-m_h^{2}}, $$ 
so that $\nh=\vec p\,/|\vec p\,|$ is a unit spatial vector, $n^\mu(\nh)$ is null, and $\omega\geq m_h$. A hadronic out state is a tensor-product state of single-hadron out states specified by these labels,
\begin{align}
	\ket{X}\;=\;\ket{\{h_r,\omega_r,\nh_r,s_r\}}\,,
	\label{eq:outstate}
\end{align}
with $r$ indexing the hadrons. $\vev{X|\psi}$, its overlap with the state in equation~\eqref{eq:state}, is the amplitude of point (iii) above.

Without loss of generality, we can introduce creation and annihilation operators for the asymptotic hadronic modes. Write $a_{h\,s}(\omega,\nh)$ for the operator annihilating an out hadron of species $h$ and spin
state $s$ with energy $\omega$ in the direction $\nh$. We will rescale it from covariant normalisation by $\sqrt{|\vec p\,|/2(2\pi)^3}$, so that
\begin{align}
	\big[a_{h\,s}(\omega,\nh),\,a^{\dagger\,s'}_{h'}(\omega',\nh')\big]_{\mp}
	\;=\;\de(\omega-\omega')\;\de^{(2)}(\nh,\nh')\;\de_{hh'}\;\de_{s}{}^{s'}\,,
	\label{eq:angalg}
\end{align}
with the upper sign for bosons and the lower for fermions. An ideal calorimeter responds additively to the hadrons that reach it; it is indifferent to hadron species; it weights each hadron by its energy; and it records the direction from which the hadron arrived (the calorimeter's location). Those four statements fix the operator:
\begin{align}
	\cE(\nh_0)\;=\;\sum_{h}\int_{m_h}^{\infty}\!\!\df\omega\int_{S^2}\!\!\df^2\nh\;\;
	\omega\;\de^{(2)}(\nh,\nh_0)\;\de^{s}{}_{s'}\;\;
	a^{\dagger\,s'}_{h}(\omega,\nh)\,a_{h\,s}(\omega,\nh)\,,
	\label{eq:Edef}
\end{align}
in which the pairing $a^{\dagger}a$ is the number operator, the sum over $h$ together with the unit matrix in spin space reflects the indifference to $h$ and $s$, the factor $\omega$ is the energy weighting, and the angular delta function reflects the localisation. This is the energy-flow operator of
refs.~\cite{Sveshnikov:1995vi,Korchemsky:1997sy,Korchemsky:1999kt,Belitsky:2001ij}, whose correlators are the energy correlators of refs.~\cite{Basham:1978bw,Basham:1978zq} (see ref.~\cite{Moult:2025nhu} for a review), and it is defined so that it projects the desired inclusive energy eigenvalue when applied to a hadronic out state
\begin{align}
	\cE(\nh_0)\ket{X}\;=\;\Big(\sum_{r}\omega_r\;\de^{(2)}(\nh_0,\nh_r)\Big)\ket{X}\,.
	\label{eq:Ediag}
\end{align}

Equation~\eqref{eq:Edef} is a distribution in $\nh_0$, so it is not yet an operator whose eigenvalue
yields a physical measurement; only its integral over a region is. What is measured is always a smeared flux, which we write as a bracket against a ``source''\footnote{We will see later why we refer to $\lambda$ as a source.} $\lambda$, e.g. for a calorimeter
\begin{align}
	\vev{\la,\cE}\;\equiv\;\int_{S^2}\!\df^2\nh\;\;\la(\nh)\,\cE(\nh)\,,
	\qquad
	\vev{\la,\cE}\ket{X}\;=\;\Big(\sum_{r}\la(\nh_r)\,\omega_r\Big)\ket{X}\,,
	\label{eq:Esmear}
\end{align}
the eigenvalue following from equation~\eqref{eq:Ediag}. All measurements performed by a calorimeter can be expressed in terms of these brackets for a suitable set of $\lambda$, e.g. taking $\la$ to be the indicator function of a calorimeter cell gives the energy in that cell.

The smeared energy flux, equation~\eqref{eq:Esmear}, is bounded by energy conservation: energy is conserved exactly, so every state $\ket{X}$ that $\ket{\psi}$ can reach satisfies
\begin{align}
	\sum_{r}\omega_r\;=\;Q\,.
	\label{eq:energycons}
\end{align}
Each hadron contributes $\la$ evaluated at its own direction times its own energy, and those
energies are positive and sum to $Q$, so
\begin{align}
	\Big|\sum_{r}\la(\nh_r)\,\omega_r\Big|\;\leq\;Q\;\max_{\nh\in S^2}\big|\la(\nh)\big|
	\label{eq:Ebound}
\end{align}
on every state the process can reach. Hence $\vev{\la,\cE}$, restricted to the fixed-energy sector relevant to $\ket{\psi}$, is a bounded self-adjoint operator and its exponentiation, $e^{-\vev{\la,\cE}}$, is defined without an ordering prescription. This will be key to later constructions.

In section~\ref{sec:PS}, we will use the detector to define a probability distribution for the joint readings of a whole array of cells, a statement of the form ``this detector records this pattern of
energies with this probability''. Quantum mechanics grants such a joint distribution if and only if the observables commute, and the smeared energy fluxes do:
\begin{align}
	\big[\,\vev{\la,\cE}\,,\,\vev{\la',\cE}\,\big]\;=\;0\,,
	\label{eq:Ecommute}
\end{align}
for any two sources, including ones that overlap.\footnote{Equation~\eqref{eq:Ecommute} is an
algebraic property of free asymptotic operators, and it is all that is used here. It should not be
confused with the field-theoretic statement that light transforms of the energy--momentum tensor
commute at non-coincident points, which is a dynamical theorem with its own conditions
\cite{Hofman:2008ar,Kravchuk:2018htv,Kologlu:2019mfz,Caron-Huot:2022eqs}. Where both definitions of
the detector exist they agree; the asymptotic one is primary here and is unconditional.}

Having defined the energy-flow operator for an idealised calorimeter, we can now ask which other out-state detectors admit joint readings. A general out-state detector that responds additively to particle states can be written in the form
\begin{align}
	\bb{D}\big[f\big]\;=\;\sum_{h}\int\!\df\omega\int\!\df^2\nh\;\;
	f_h{}^{s}{}_{s'}(\omega,\nh)\;\;a^{\dagger\,s'}_{h}(\omega,\nh)\,a_{h\,s}(\omega,\nh)
	\label{eq:D1}
\end{align}
for some kernel $f$ which describes the detector response, e.g. $\bb{D}[\omega\lambda] = \vev{\la,\cE}$. By applying equation~\eqref{eq:angalg} twice,
\begin{align}
	\big[\bb{D}[f],\,\bb{D}[g]\big]\;=\;\bb{D}\big[\,[f,g]\,\big]\,,
	\label{eq:algebra}
\end{align}
with the matrix commutator taken pointwise. Hence
\begin{align}
	\big[\bb{D}[f],\,\bb{D}[g]\big]\;=\;0
	\qquad\text{iff}\qquad
	\big[f,g\big](\omega,\nh)\;=\;0\quad\text{for every }\omega,\nh\,.
	\label{eq:commutecriterion}
\end{align}
We note that $[\bb{D}[\omega\lambda] , \bb{D}[f]]=0$ for all $f$. More generally, $[\bb{D}[f],\,\bb{D}[g]]$ vanishes for any set of kernels diagonal in one fixed spin basis, a necessary ingredient for section~\ref{sec:EG}. Additionally, for any $f_h{}^{s}{}_{s'} = \bar f_h \delta^{s}{}_{s'}$, 
$$\bb{D}\big[\bar f_h\, \delta^{s}{}_{s'} \big] \ket{X} = \sum_r \bar f_{h_r}(\omega_r,\nh_r)\ket{X} .$$ 

Kernels with off-diagonal entries do not, in general, commute. Consequently, a helicity flip gives an operator with no joint probabilistic reading at all and will suffer an inherent uncertainty relation.\footnote{Indeed, if one instead formulated the ``detectors'' perturbatively such that the operator algebra carries colour, and so colour correlations could be defined, then a detector implementing a colour rotation would likewise admit no joint probabilistic interpretation.}  An off-diagonal Hermitian kernel still defines an observable, but it cannot in general be included in the same jointly measurable detector family. Every family of detector operators used in this work is deliberately kept inside the commuting class since this is the class typically measured by collider detectors.

Finally, we will assume the out states of QCD are spanned by states of stable free hadrons, as is standard in the usual $S$-matrix formulation,
\begin{align}
	\id\;&=\;\sum_{N}\frac{1}{N!}\sum_{\{h_r,s_r\}}\int\prod_{r=1}^{N}\df\Pi_{h_r}\;\;\ket{X}\bra{X}
	\;\equiv\;\sumint_{X}\ket{X}\bra{X}\,,
	\nn\\[4pt]
	\df\Pi_{h}\;&=\;\frac{\sqrt{\omega^{2}-m_h^{2}}\;\;\df\omega\,\df^2\nh}{2(2\pi)^3}\,,
	\qquad \omega\;\geq\;m_h\,,
	\label{eq:completeness}
\end{align}
so that the total cross section is
\begin{align}
	\sigma_{\rm tot}\;=\;\sumint_{X}\big|\vev{X|\psi}\big|^{2}\,.
	\label{eq:sigmatot}
\end{align}

\subsection{Measurements across multiple calorimeter cells}
\label{sec:PS}

We can now build the parton-shower functional measure, the object measured by a simultaneous array of calorimeter cells. Let us start by considering a simple physical detector, made up of a uniform lattice of calorimeter cells.
Segment the celestial sphere into patches $\Delta S_1,\dots,\Delta S_m$, one per calorimeter
element, with indicator functions 
\begin{align}
    \chi_a(\hat n) =  \left\{
        \begin{matrix}
            1 ~ {\rm for} ~ \hat n \in \Delta S_a, \\
            0 ~ {\rm otherwise},
        \end{matrix} \right.
\end{align}
so that
\begin{align}
	\vev{\chi_a,\cE}\;=\;\int_{\Delta S_a}\!\!\df^2\nh\;\;\cE(\nh)
	\label{eq:cellop}
\end{align}
is the operator reading the energy in cell $a$. At the end of an event the detector reports $m$
numbers, one per cell, and its complete output is the joint distribution of those numbers. Since the
readings commute, that distribution exists, and it is given by
\begin{align}
	\df E_m\;=\;\bra{\psi}\;\prod_{a=1}^{m}\df E_a\;\;\de\Big(\vev{\chi_a,\cE}-E_a\Big)\;\ket{\psi}\,.
	\label{eq:dEm}
\end{align}
Equation~\eqref{eq:dEm} contains every energy measurement this detector can make, because any such measurement is some function of the $E_a$ and is obtained by integrating $\df E_m$ against that function. It also contains no redundant information because every variable in it is a number the detector actually reports. Furthermore, it is a positive measure of total weight $\sigma_{\rm tot}(Q)$, supported on the region where every reading is non-negative and the readings sum to $Q$, which is to say a probability distribution over what the detector can see (once normalised).

Equation~\eqref{eq:dEm} depends on the detector segmentation, and that is a property of the apparatus rather than of the theory. To remove it, we refine the detector cells. Merging two cells gives
$\chi_{a\cup b}=\chi_a+\chi_b$, so
\begin{align}
	\vev{\chi_{a\cup b},\cE}\;=\;\vev{\chi_a,\cE}+\vev{\chi_b,\cE}
	\label{eq:additive}
\end{align}
trivially. The distribution for a coarser segmentation is therefore recovered
from that for any refinement by adding the corresponding cell readings.

We can take the limit that the size of the detector cells goes to zero, in which case the natural quantity is the energy per unit solid angle,
\begin{align}
	E(\nh_a)\;\equiv\;\frac{E_a}{\Delta S_a}\,,
	\label{eq:fluxvar}
\end{align}
which has the interpretation of an angular energy flux. If the flux were a smooth field on the
sphere then, for a sufficiently small cell,
\begin{align}
	\int_{\Delta S_a}\!\!\df^2\nh\;\;\cE(\nh)\;\ra\;\Delta S_a\;\cE(\nh_a)\,,
	\label{eq:smoothstep}
\end{align}
so each factor of equation~\eqref{eq:dEm} would become
\begin{align}
	\df E_a\;\;\de\Big(\vev{\chi_a,\cE}-E_a\Big)
	\;\ra\;
	\frac{\df E_a}{\Delta S_a}\;\;\de\Big(\cE(\nh_a)-\frac{E_a}{\Delta S_a}\Big)
	\;=\;\df E(\nh_a)\;\;\de\Big(\cE(\nh_a)-E(\nh_a)\Big)\,.
	\label{eq:cellrescale}
\end{align}
Equation~\eqref{eq:dEm} is a product over cells, and so the continuum limit must be taken on the
product rather than on a single factor. The product of the measures defines a functional measure,
and the product of the delta functions defines a functional delta,
\begin{align}
	\lim_{m \rightarrow \infty } \prod_{a=1}^{m} \df E(\nh_a)\;\ra\;[\cD E]\,,
	\qquad
	\lim_{m \rightarrow \infty}\prod_{a=1}^{m} \de\Big(\cE(\nh_a)-E(\nh_a)\Big)\;\ra\;\de\big[\cE-E\big]\,,
	\label{eq:continuumreplace}
\end{align}
the latter imposing one constraint at every point of the celestial sphere rather than one constraint
in total. The functional delta is defined by its action on a test functional,
\begin{align}
	\int\!\big[\cD E\big]\;\;\de\big[\cE-E\big]\;\;\Phi\big[E\big]\;=\;\Phi\big[\cE\big]\, .
	\label{eq:funcdeltadef}
\end{align}
With this, we arrive at the \emph{parton shower measure},
\begin{align}
	\lim_{m\rightarrow \infty}\df E_m = \cD\PS[E]\;\cong \;[\cD E]\;\;\bra{\psi}\;\de\big[\cE-E\big]\;\ket{\psi}\,,
	\label{eq:PSformal}
\end{align}
the probability density functional that the detector records the flux $E$. For a calorimeter detector an event is an energy flux on the celestial sphere, and equation~\eqref{eq:PSformal} is the probability density functional measure for how often the theory produces that event.

This argument identifies the correct continuum object, but its intermediate interpretation requires qualification: one step in that derivation lacked sufficient care and hence equation~\eqref{eq:PSformal} is given heuristically, marked with a $\cong$. Equation~\eqref{eq:smoothstep} is legitimate only if the flux is a smooth field on the sphere, so that halving a cell halves its reading. This may be valid in some idealised scenarios, in which case equation~\eqref{eq:PSformal} is meant literally. However, this is not the case in QCD.

In QCD the energy flux carried by an individual hadronic state is not a smooth field, since the theory has a mass gap. Every state reached from $\ket{\psi}$ carries total energy $Q$ and every hadron carries at least $m_\pi$, so no configuration contains more than $\lfloor Q/m_\pi\rfloor$ hadrons. Once the calorimeter has more cells than that, most cells read exactly zero; and shrinking a cell no longer reduces its reading in proportion to its area, since below a certain size a cell either contains a hadron or it does not. The flux realised on a hadronic final state, $\ket{X},$ is not a smooth field but a distribution describing a finite set of hadrons on the sphere,
\begin{align}
	E_X(\hat n)\;=\;\sum_{r}\omega_r\,\delta^{(2)}(\nh_r , \nh)\,,
	\qquad\text{so that}\qquad
	\vev{\la,E_X}\;=\;\sum_{r}\omega_r\,\la(\nh_r)\,.
	\label{eq:fluxconfig}
\end{align}
As it is a distribution, it has meaning only when integrated against something.

If $E$ is distributional then so is any functional of it, and $\PS[E]$,
a density on flux profiles, is not a number that can be quoted at a given $E$. The measure, however, is well defined, which is why equation~\eqref{eq:PSformal} is written as $\cD\PS[E]$ and never simply as $[\cD E]\PS[E]$. The complete definition of $\cD\PS[E]$ is given in appendix~\ref{app:atomic}; however, its full formulation, including the subtleties associated with its distributional nature, is not necessary for the present.
Rather, the measure can be rigorously defined through its action on a test functional (measurement),
\begin{align}
	\int\!\cD\PS[E]\;\;\Phi\Big(\vev{\la_1,E},\dots,\vev{\la_M,E}\Big)
	\;=\;\bra{\psi}\;\Phi\Big(\vev{\la_1,\cE},\dots,\vev{\la_M,\cE}\Big)\;\ket{\psi}\,,
	\label{eq:PSdef}
\end{align}
for any sources $\la_1,\dots,\la_M$ and any functional $\Phi$ of the corresponding readings. A real
detector has cells of finite size and reports finitely many numbers, so equation~\eqref{eq:PSdef} covers
every measurement that could be made (consider setting $\lambda_a = \chi_a$). However, its distributional nature limits the ability to speak of the flux at a point, which is not a physical measurement. We keep equation~\eqref{eq:PSformal} as an informal picture to carry, while equation~\eqref{eq:PSdef} provides the precise meaning.

The dynamical content of equation~\eqref{eq:PSdef} can be exposed by the insertion of two identities on the right-hand side. Place
completeness relations, equation~\eqref{eq:completeness}, on both sides of $\Phi$. Since $\cE$ is diagonal on out states by equation~\eqref{eq:Ediag}, so is any function of it, and the double sum collapses to a single one, leaving
\begin{align}
	\int\!\cD\PS[E]\;\;\Phi\big(\vev{\la_1,E},\dots\big)
	\;=\;\sumint_{X}\big|\vev{X|\psi}\big|^{2}\;\;\Phi\big(\vev{\la_1,E_X},\dots\big)\,.
	\label{eq:PSpush}
\end{align}
The parton shower measure is the hadronic $S$-matrix matched onto flux configurations. States that produce the same flux configuration are degenerate and so are integrated out. The two identity insertions collapsed because the kernel of $\cE$ has no off-diagonal spin structure, the same restriction that made the detectors commute; with an off-diagonal kernel a double sum over states differing only in their internal labels survives, and equation~\eqref{eq:PSpush} becomes more complicated.

Setting $\Phi=1$ in equation~\eqref{eq:PSpush} gives the total weight $\sigma_{\rm tot}(Q)$, and the right-hand side is manifestly non-negative for any non-negative measurement. These two facts ensure $\cD\PS$ is a sampleable
distribution and not just a generating functional: divided by $\sigma_{\rm tot}$ it is a probability distribution over flux configurations. To say that a parton shower ``computes'' $\cD\PS$ is to say that it samples this probability distribution.

\subsection{Laplace space parton shower functional}
\label{sec:laplace}

In the previous section we defined a measure against which any calorimeter measurement can be computed. However, the measure is awkward to work with since each statement about it requires a choice of test function. This, however, is a familiar situation and has a familiar remedy. We introduce an invertible integral transformation, and so look at a lossless encoding of the information in a friendlier representation. Take the test function to be an exponential of the pairing, and define
\begin{align}
	\PSL[\la]\;\equiv\;\int\!\cD\PS[E]\;\;e^{-\vev{\la,E}}
	\;=\;\bra{\psi}\;e^{-\vev{\la,\cE}}\;\ket{\psi}\,.
	\label{eq:PSL}
\end{align}
This is a functional generalisation of a Laplace transform. The only difference from the usual one
is that the conjugate variable is not a number but a function $\la(\nh)$ on the sphere, one value
for each direction in which the flux can be measured. We work with the
transform instead of the distribution since $\PSL[\la]$ is a number
for every bounded source $\la$. We refer to $\PSL[\la]$ as the \emph{parton shower source functional}, or simply the \emph{parton shower functional}, and it is one of our two core objects.

$\la(\hat n)$ also carries a physical interpretation. Necessarily, the eigenvalue of $\vev{\la,\cE}$ in equation~\eqref{eq:Esmear} is dimensionless and so $\la$ carries mass dimension $-1$. Consequently,
\begin{align}
	\mu(\nh)\;\equiv\;\frac{1}{\la(\nh)}
	\label{eq:resolution}
\end{align}
is an energy for real and positive $\lambda$. A hadron of energy $\omega$ arriving at $\nh$ is weighted in
equation~\eqref{eq:PSL} by $e^{-\omega/\mu(\nh)}$, so $\mu(\nh)$ is a \emph{resolution scale}: the
measurement encoded by $\la$ is blind to quanta far below $\mu(\nh)$ since $e^{-\omega/\mu(\nh)} \rightarrow 1$, and exponentially suppresses those far above $\mu(\nh)$. Only hadrons arriving at $\nh$ for which $\omega \sim \mu(\nh)$ contribute; though, as we will see in section~\ref{sec:oneloop}, the actual scales entering logarithms are a little more complicated. $\mu(\nh)$ is also a field over the celestial sphere and so there is one such scale at every point.

Putting completeness relations on both sides of the exponential in equation~\eqref{eq:PSL}, and using that $\cE$ is diagonal on out states,
\begin{align}
	\PSL[\la]\;=\;\sumint_{X}\;\big|\vev{X|\psi}\big|^{2}\;\prod_{r}e^{-\la(\nh_r)\,\omega_r} \, .
	\label{eq:PSLproduct}
\end{align}
The parton shower functional is the hadronic cross section ($S$-matrix), reweighted hadron by hadron by
\begin{align}
	w(\omega,\nh)\;=\;e^{-\la(\nh)\,\omega}\;=\;e^{-\omega/\mu(\nh)}\,,
	\qquad 0<w(\omega,\nh)\leq1\,  .
	\label{eq:weight}
\end{align}

$\PSL$ is infrared and collinear safe. It is defined in terms of the infrared- and collinear-safe energy-flow operator and the exponentiation of the operator has not broken the safety. More explicitly, a soft quantum of energy $\omega\ra0$ multiplies the weight by $e^{-\la(\nh)\omega}\ra1$ and so the measurement is soft safe. A parent of energy $\omega$ splitting into daughters
of energies $z\omega$ and $(1-z)\omega$ multiplies it by
\begin{align}
	e^{-\la(\nh_1)z\omega}\;e^{-\la(\nh_2)(1-z)\omega}
	\;\xrightarrow[\ \nh_1,\nh_2\ra\nh\ ]{}\;e^{-\la(\nh)\omega}\,,
	\label{eq:collinear}
\end{align}
and so it is also collinear safe. Two ingredients are responsible for this. The first is that energy is additive. The second is that the kernel is a spin
scalar summed over species, so that a splitting which changes species or spin does not move flux from one spin detector to another.

\subsection{Computing Observables in Laplace Space}

Let us now give some simple examples of computing observables. Differentiating equation~\eqref{eq:PSL} with respect to the source brings down powers of the flux, and setting the source to zero afterwards leaves the energy
correlators,
\begin{align}
	\frac{\de^{k}\,\PSL[\la]}{\de\big(\!-\!\la(\nh_1)\big)\cdots\de\big(\!-\!\la(\nh_k)\big)}
	\bigg|_{\la=0}
	\;=\;\bra{\psi}\,\cE(\nh_1)\cdots\cE(\nh_k)\,\ket{\psi}\, .
	\label{eq:correlators}
\end{align}
These are the moments of $\cD\PS$, and they fully determine it. Every configuration carries the same total energy $Q$, each pairing $\vev{\la,E}$ is confined to a bounded range, and a distribution on a bounded range is fixed by its moments, which is to say the Bromwich contour is well defined (more on this at the end of this subsection). The infinite set of energy correlators is therefore also a complete characterisation of the energy flux. 

Event shapes come instead from fixing the angular profile of the source and varying only its overall scale: for a fixed weight function $f(\hat n)$ and the observable
$\tau=\vev{f,E}$, set $\la(\hat n)=\nu f(\hat n)$, so that
\begin{align}
	\int\!\df\tau\;e^{-\nu\tau}\;\frac{\df\sigma}{\df\tau} \;=\; \int\!\df\tau\;e^{-\nu\tau} \int \cD {\rm PS}[E] ~ \delta\left(\tau - \vev{f,E}\right) \;=\;\PSL\big[\la\big]\Big|_{\la = \nu f} ~ .
	\label{eq:eventshape}
\end{align}
Hence, an event shape in Laplace space, equation~\eqref{eq:eventshape}, is just equation~\eqref{eq:PSL} with $\la$ restricted to the one-parameter family $\nu f$. A Laplace-space event-shape resummation formula is therefore nothing other than $\PSL$ evaluated on such a family \cite{Korchemsky:1999kt,Belitsky:2001ij}, and this is the cheapest check available on any candidate expression for $\PSL$: put in the thrust weight and compare against what is known, we do this in section~\ref{sec:thrust}.

$\PSL$'s positivity is inherited from the parton shower measure. Each derivative in equation~\eqref{eq:correlators} brings down a
factor of the flux, which is non-negative, so
\begin{align}
	(-1)^{k}\,\frac{\de^{k}\,\PSL[\la]}{\de\la(\nh_1)\cdots\de\la(\nh_k)}\;\geq\;0
	\qquad\text{for all }\la\geq0\,,
	\label{eq:cm}
\end{align}
and in particular every energy correlator is non-negative. Positivity further requires that $\PSL$ is bounded and decreasing,
\begin{align}
	0 = \PSL[\infty] \;<\;\PSL[\la]\;\leq\;\PSL[0]\;=\;\sigma_{\rm tot}(Q)\qquad\text{for all }\la\geq0\,.
	\label{eq:bounds}
\end{align}
The analytic structure of $\PSL$ follows from energy conservation. Every configuration carries total energy $Q$, so for any smooth and bounded $\lambda(\hat n) \in \mathbb{C}$ the integrand of
equation~\eqref{eq:PSL} obeys
\begin{align}
	\big|e^{-\vev{\la,E}}\big|\;=\;e^{-\vev{\Re\la,\,E}}\;\leq\;
	\exp\Big[\,Q\;\max_{\nh\in S^2}\big|\Re\la(\nh)\big|\,\Big]\,,
	\label{eq:complexbound}
\end{align}
and so the integral converges everywhere, and
\begin{align}
	\big|\PSL[\la]\big|\;\leq\;\sigma_{\rm tot}(Q)\;
	\exp\Big[\,Q\;\max_{\nh\in S^2}\big|\Re\la(\nh)\big|\,\Big]\, ,
	\label{eq:entirebound}
\end{align}
ensuring $\PSL$ is finite. Infrared divergences, Landau poles and renormalon factorial divergences appear routinely in intermediate perturbative expressions, but none are an intrinsic non-perturbative property of $\PSL$.

This analyticity ensures that singularities encountered in a particular perturbative representation are not singularities of the exact parton shower functional, and it permits distributional inversion on finite-dimensional source families. Nothing was lost by moving from working with $\cD\PS$ to $\PSL$. Whilst the restriction to an exponential of smeared fluxes can look like a loss of information, it is not. 

We close this section by providing the recipe to compute an entirely generic calorimeter observable. A detector has finitely many cells, and anything computed from its output is a function of finitely many readings, so let $f_1,\dots,f_M$ be the weight functions that a given measurement resolves and let $\tau_i=\vev{f_i,E}$ be the corresponding
readings. Restricting the source to the span of those weights turns equation~\eqref{eq:PSL} into an ordinary
$M$-dimensional Laplace transform of their joint distribution,
\begin{align}
	\int\!\df^{M}\tau\;\;e^{-\sum_i\nu_i\tau_i}\;\;
	\frac{\df\sigma}{\df\tau_1\cdots\df\tau_M}
	\;=\;\PSL\Big[\la=\sum_{i=1}^{M}\nu_i\,f_i\Big]\,,
	\label{eq:jointtransform}
\end{align}
of which the event shape in equation~\eqref{eq:eventshape} is the case $M=1$ and $f_1$ is continuous over the entire celestial sphere. By equation~\eqref{eq:entirebound} the right-hand side is entire in every $\nu_i$, so the joint distribution comes back from an $M$-fold
Bromwich contour,
\begin{align}
	\frac{\df\sigma}{\df\tau_1\cdots\df\tau_M}
	\;=\;\prod_{i=1}^{M}\bigg[\,\frac{1}{2\pi i}\int_{c_i-i\infty}^{c_i+i\infty}\!\!\df\nu_i\;\;
	 e^{\nu_i\tau_i}\,\bigg]\;\;\PSL\Big[\la=\sum_{j}\nu_j\,f_j\Big]\,,
	\label{eq:bromwich}
\end{align}
and any observable built from those readings, $v=u(\tau_1,\dots,\tau_M)$, follows from it by
\begin{align}
	\frac{\df\sigma}{\df v}\;=\;\int\!\df^{M}\tau\;\;
	\de\big(v-u(\tau_1,\dots,\tau_M)\big)\;\;
	\frac{\df\sigma}{\df\tau_1\cdots\df\tau_M}\,.
	\label{eq:genericobs}
\end{align}
The recipe is therefore to write the observable in terms of the smeared fluxes on which it depends, restrict the source to the span of the corresponding weights, invert, and integrate against the observable's definition. The energy correlators and the event shapes above are the two simplest cases. What lies outside the reach of $\PSL$ is an observable that depends on more than calorimetry: one that asks which particle carried the
energy, or how many there were. This is the focus of the next subsection.

\subsection{Beyond calorimetry}
\label{sec:EG}

$\PSL$ defines measurements made only by calorimetry. That is sufficient for typical infrared- and collinear-safe measurements but leaves a great deal of what is measured unaddressed: multiplicities, identified-particle spectra, anything defined on tracks. An observable of that kind resolves more detail from the final state. We must now resolve which hadrons arrive, in which spin states, and how many of them there are. 

The detector family must be enlarged without leaving the commuting class, following the construction in section~\ref{sec:detectors}. We therefore introduce a more general detector: for a species $h$, a general
$\DLz\leq 0$ and a spin matrix $\Sigma$ (diagonal in the fixed basis), let
\begin{align}
	\bb{D}^{[\DLz]}_{h,\Sigma}(\nh_0)\;=\;\int_{m_h}^{\infty}\!\!\df\omega\int_{S^2}\!\!\df^2\nh\;\;
	\omega^{\,-\DLz}\;\de^{(2)}(\nh,\nh_0)\;\Sigma^{s}{}_{s'}\;\;
	a^{\dagger\,s'}_{h}(\omega,\nh)\,a_{h\,s}(\omega,\nh)\,.
	\label{eq:D3}
\end{align}
The species label $h$ and the matrix $\Sigma$ determine which hadrons in which spin states the detector
responds to; the power $\omega^{-\DLz}$ weights them, so that $\DLz=0$ simply counts them, $\DLz=-1$ weights them by energy, $\DLz=-2$ by energy squared, and so on. By construction the energy-flow detector is a member of the extended family,
\begin{align}
	\cE(\nh_0)\;=\;\sum_{h}\bb{D}^{[-1]}_{h,\id}(\nh_0) \, .
	\label{eq:hinge}
\end{align}
Through their sensitivity to hadronic quantum numbers, $\bb{D}^{[\DLz]}_{h,\Sigma}$ are operator statements of what tracking detectors measure, and their perturbative description is known to require non-perturbative track or fragmentation functions
\cite{Chen:2020vvp,Chang:2025zib,Gonzalez:2026aha}.

We refer to the index $\DLz$ as the \emph{canonical (free) light-ray dimension} of the detector. With the
mode normalisation of equation~\eqref{eq:angalg} the operator in equation~\eqref{eq:D3} carries mass dimension $-\DLz$. 
In the massless limit, i.e. neglecting corrections which scale as $m_h^{2}/\omega^{2}$, $\DLz$ has a familiar interpretation.

In this limit, the generalised detectors become light-ray operators, which are well studied in QCD and CFTs \cite{Sveshnikov:1995vi,Korchemsky:1997sy,Korchemsky:1999kt,Belitsky:2001ij,Hofman:2008ar,Caron-Huot:2022eqs,Chang:2025zib,Chang:2025kgq,Gonzalez:2026aha}. In the literature, light-ray operators are more commonly specified by their Lorentz eigenvalues ($J_L$, $\Delta_L$, and $J$ for local operators). These are determined by $\DLz$ as we will now discuss. The weight of a light-ray operator under a collinear boost is given by its light-ray spin $J_L$, defined by
\begin{align}
	\Lambda_\eta : \bb{D}^{[\DLz]}_{h,\Sigma}(\nh)\;\mapsto \;\eta^{\,J_L}\;\bb{D}^{[\DLz]}_{h,\Sigma}(\nh)
	\qquad\text{under}\qquad \Lambda_\eta :
	(1,\nh)\;\mapsto\;\eta\,(1,\nh)\,,
	\label{eq:boostweight}
\end{align}
and, for the so-called \emph{massless twist-2 DGLAP detectors} \cite{Caron-Huot:2022eqs,Chang:2025zib,Gonzalez:2026aha}, the class of light-ray operators most relevant to our construction,
\begin{align}
	\DLz\;=\;J_L + d - 2\,,
	\label{eq:JzeroJL}
\end{align}
with $d$ the spacetime dimension. Additionally, for bare operators that are the light transform of a local primary of spin $J$, we have $\bb{D}^{[- \DLz + 1= J]}_{h,\Sigma}=\mathbf{L}\big[O^{[J]}_{h,\Sigma}\big]$. For instance, the energy flux $\cE$ is the light transform of the energy--momentum tensor which is a spin-2 local operator $O^{[2]} = T$. The dimension $\DLz$ is protected by symmetry through its relation to $J_L$, and so is preserved under renormalisation, whilst $J$ is not. Consequently, in renormalised perturbation theory
\begin{align}
	J\;= - \DLz + 1 - 2\ga^{T}(- \DLz + 1)\,,
	\label{eq:Janom}
\end{align}
with $\ga^{T}$ the time-like DGLAP anomalous dimension \cite{Chen:2020uvt}. Thus $J$ runs, while $\DLz$ does not\footnote{We emphasise that $\DLz$ is fixed by the operator's mass dimension and collinear spin; it is distinct from the light-ray dimension $\Delta_L$ \cite{Kravchuk:2018htv,Kologlu:2019mfz} which also runs. Their values coincide in the free-theory limit.}. We label the family of detectors by $\DLz$ throughout since it is most natural when applied to detectors for QCD with a mass gap. This label is fixed by the detector definition alone, and so is robust whether or not the operator descends from a local primary, which a general light-ray operator need not do.

We must now construct a functional for the measurement of fluxes determined by $\bb{D}^{[\DLz]}_{h,\Sigma}(\nh_0)$. Everything now proceeds as before, since nothing in sections~\ref{sec:PS} and \ref{sec:laplace} used any property of $\cE$ beyond its membership within the commuting class, its boundedness, and its diagonality on out states. Each $\bb{D}^{[\DLz]}_{h,\Sigma}$ shares these properties, with boundedness following from the mass gap and energy conservation. The event generator measure is 
\begin{align}
	\cD\EG[\{E\}]\;\cong \;\;\bra{\psi}\;\prod_{\DLz, h, \Sigma}\left[\cD E^{[\DLz]}_{h,\Sigma}\right]\;\de\left[\bb{D}^{[\DLz]}_{h,\Sigma}-E^{[\DLz]}_{h,\Sigma}\right]\;\ket{\psi}\,,
	\label{eq:EGformal}
\end{align}
and we move immediately to the Laplace space functional. The extended family of detectors is now indexed by three labels, so we share these labels with each source and define
\begin{align}
	\EGL\big[\{\la\}\big]\;\equiv\;\bra{\psi}\;\exp\Big[-\!\!\sum_{\DLz,h,\Sigma}
	\big\langle\la^{[\DLz]}_{h,\Sigma},\,\bb{D}^{[\DLz]}_{h,\Sigma}\big\rangle\Big]\;\ket{\psi}\, ,
	\label{eq:EGL}
\end{align}
the \emph{event generator source functional}, or simply the \emph{event generator functional}. 

Although the exponent of equation~\eqref{eq:EGL} carries a source for every detector, the whole sum can also be understood as a single operator of the form given in equation~\eqref{eq:D1}: the sources may be collected into one kernel
$\ell$, giving
\begin{align}
	\sum_{\DLz , h,\Sigma}\big\langle\la^{[\DLz]}_{h,\Sigma},\,\bb{D}^{[\DLz]}_{h,\Sigma}\big\rangle
	\;=\;\bb{D}\big[\ell\big]\,,
	\qquad
	\ell_h{}^{s}{}_{s'}(\omega,\nh)\;=\;\sum_{\DLz}\ \sum_{\Sigma}
	\la^{[\DLz]}_{h,\Sigma}(\nh)\;\Sigma^{s}{}_{s'}\;\omega^{-\DLz}\, ,
	\label{eq:EGexponent}
\end{align}
and, by the diagonality of $\Sigma$, this combined operator is diagonal on out states,
\begin{align}
	\bb{D}\big[\ell\big]\,\ket{X}\;=\;
	\Big(\sum_{r}\ell_{h_r}{}^{s_r}{}_{s_r}(\omega_r,\nh_r)\Big)\,\ket{X}\,,
	\qquad\text{no sum on }s_r\,.
	\label{eq:EGexponenteig}
\end{align}
The diagonal entry $\ell_h{}^{s}{}_{s}(\omega,\nh)$ is therefore what a single hadron of species $h$ in spin state $s$, arriving with energy $\omega$ at $\nh$, contributes to the exponent in equation~\eqref{eq:EGL}.
Explicitly, inserting a final-state completeness relation gives an expression in terms of the $S$-matrix,
\begin{align}
	\EGL\big[\{\la\}\big]\;=\;\sumint_{X}\big|\vev{X|\psi}\big|^{2}\;\prod_{r}
	w\big(\omega_r,\nh_r,h_r,s_r\big)\,,
	\label{eq:EGLproduct}
\end{align}
where,
\begin{align}
	w\big(\omega,\nh,h,s\big)\;=\;e^{-\ell_h{}^{s}{}_{s}(\omega,\nh)}
	\;=\;
	\exp\Big[-\!\!\sum_{\DLz}\ \sum_{\Sigma}
	\la^{[\DLz]}_{h,\Sigma}(\nh)\;\Sigma^{s}{}_{s}\;\omega^{-\DLz}\Big]\,,
	\qquad\text{no sum on }s\,.
	\label{eq:EGweight}
\end{align}
Here, the sum over $s$ is a part of the sum over $X$ in equation~\eqref{eq:EGLproduct}, see equation~\eqref{eq:outstate}. Comparing equation~\eqref{eq:EGweight} with
equation~\eqref{eq:weight}, the added dependence on $h$, $s$ and $\DLz$ is the entire difference between the
two functionals. The exponent is bounded, and is positive whenever the sources are, just as for $\PSL$. Consequently, $\EGL$ is positive with total weight $\sigma_{\rm tot}$, encoding the entirety of $\cD\EG$, which has an interpretation as a probability distribution. 

$\cD\PS$ is the marginal of $\cD\EG$ in which everything but the total energy flux has been integrated out. In Laplace space, marginalising is simply switching off the sources one does not want:
\begin{align}
	\PSL[\la]\;=\;\EGL\big[\{\la\}\big] \Big|_{\la^{[\DLz]}_{h,\Sigma}=\la\;\de_{\DLz,-1}\;\de_{\Sigma,\id}}\, .
	\label{eq:restriction}
\end{align}
With this restriction, equation~\eqref{eq:EGexponent} collapses to $\ell_h=\la(\nh)\,\omega\,\id$, and the
exponent of equation~\eqref{eq:EGL} to $\vev{\la,\cE}$. 

The price for the generality of $\EGL$ is the loss of explicit infrared and collinear safety. Two properties of $\cE$ made $\PSL$ infrared and
collinear safe: the weight was the first power of the energy, and the detector was blind to species
and spin. Repeating the collinear test on equation~\eqref{eq:EGweight}: a splitting sends
\begin{align}
	\omega^{-\DLz}\;\ra\;(z\omega)^{-\DLz}+\big((1-z)\omega\big)^{-\DLz}
	\;=\;\omega^{-\DLz}\Big[z^{-\DLz}+(1-z)^{-\DLz}\Big]\,,
	\label{eq:splitweight}
\end{align}
and so for $z\in(0,1)$
\begin{align}
	z^{-\DLz}+(1-z)^{-\DLz}\;
	\begin{cases}
		=1\,, & \DLz=-1\,,\\
		>1\,, & \DLz > -1\,,\\
		<1\,, & \DLz < -1\,.
	\end{cases}
	\label{eq:additivity}
\end{align}
The detector is collinear safe only at $\DLz=-1$. For $\DLz\neq -1$ the detector is collinear unsafe, and for $\DLz = 0$ soft sensitive as well. The energy weight $\DLz=-1$, summed over species and over spin (that is $\cE$), is the only intrinsically infrared- and collinear-safe member of the family. Every other member fundamentally requires non-perturbative input and so becomes perturbatively safe only when careful cancellation is manifest \cite{Collins:1981uw,Collins:2011zzd,Chen:2020vvp,Lee:2023tkr,Jaarsma:2023ell}. Nevertheless a generator must resolve the unsafe directions.

$\EGL$ provides a complete description of any physical event. The architecture of a detector will determine which $\lambda^{[\DLz]}_{h, \Sigma}(\hat n)$ are ``switched on'', i.e. take non-zero values.  However, each $\bb{D}^{[\DLz]}_{h,\Sigma}$ reports only a single Mellin moment of a given hadronic species and spin state, with the energy ranging over the compact interval $m_h \leq \omega \leq Q$. A complete tower of readings in $\DLz$ is therefore required to fully reconstruct the energy dependence of the hadrons. There is nonetheless some freedom in the choice of tower: $-\DLz \in \mathbb{N}$ is the conventional choice for Mellin inversion, but other infinite sets may also suffice.

\subsection{Remarks}

To summarise: the parton shower functional $\PSL$ and the event generator functional $\EGL$ are matrix elements of bounded positive operators built from asymptotic modes. Both are exact, hadron-level and non-perturbative; both evaluate to the hadronic cross
section reweighted by independent single-hadron factors; and both define positive normalised distributions whose draws are events. $\PSL$ is a marginal of $\EGL$, the energy flux is the unique safe detector in the family, and the resolved readings contain the entire measurable final state. The challenge now lies in computing these functionals, a task to which we turn in the following sections.

The role of these functionals is twofold. First, and most straightforwardly, they provide a precise answer to the question of generator accuracy by defining exactly what generators must compute. The functionals are physical observables and, when computed perturbatively, admit both an expansion in $\as$ and a logarithmic expansion. $\PSL$ encodes all observables accessible to a calorimetric detector, and thus validating a generator against $\PSL$ validates it for all calorimetric measurements. Likewise, validation against $\EGL$ provides a definitive test of generator accuracy. We will explore $\PSL$ at one loop in the next section; however, the broader question of how these functionals should be used to assess parton-shower accuracy is left to future work.

The second, and substantially more ambitious, role of these functionals is as a framework for event-generator construction. The functionals are defined non-perturbatively. A robust factorisation theorem for their computation would provide an analytically controlled formulation of event generation, with parton showering emerging from the perturbative evolution of the factorised components, and hadronisation encoded in the low-scale matrix elements. The inclusive matching of low-scale hadronic detector operators onto partonic operators has recently been understood in terms of generalised fragmentation \cite{Chang:2025zib,Lee:2025okn}, suggesting a direct connection between hadronisation in event generators and generalised fragmentation functions. We leave the exploration of this direction to future work.

\section{Computing $\PSL$ at one loop in $e^+ e^-$}
\label{sec:oneloop}

In the previous section we defined the parton shower functional non-perturbatively. Now we pass to perturbation theory to compute the functional at one loop for $\gamma^* \rightarrow {\rm partons}$ in massless perturbative QCD. The object we are computing is
\begin{align}
	\PSL[\la]\;=\;\sum_{p}\sumint_{Y \in \{q \bar q , q \bar q g\}}\;\big|\bra{Y}\eps^{p}_\mu(q)\int\df^4x\;e^{iq\cdot x}\,J^\mu(x) \ket{0} \big|^{2}\;\prod_{i \in Y}e^{-\la(\nh_i)\,\omega_i} \, ,
	\label{eq:PSLproductpert}
\end{align}
where $J$ is the quark electromagnetic current and $p$ the polarisation of the virtual photon. Our goal is to compute only the logarithmic terms relevant to parton-shower construction.

Let us begin by laying out the kinematics and properties of $\PSL$. Firstly, we label the final-state partons with momenta $p_i \in \{p_q , p_{\bar q}, p_g\}$. The energy flux defined by an $m$-parton final state is given by
\begin{align}
    E_{m}(\hat n) = \sum^{m}_{i=1} \omega_i ~ \delta^{(2)}(\hat n_i , \hat n)\, ,
\end{align}
so that 
\begin{align}
    \vev{\la,E_m} = \sum_{i=1}^{m} \lambda(\hat n_i) \omega_i \, .
\end{align}
Next, consider letting $\lambda'(\hat n) = \lambda(\hat n) + v_{\mu} n^{\mu}$. In this case,
\begin{align}
    \vev{\la',E_m} = \vev{\la,E_m} + v \cdot q \, ,
\end{align}
for any choice of $v$, and as a result,
\begin{align}
    \PSL[\lambda'] = e^{- v \cdot q} ~ \PSL[\lambda] \, , \label{eq:moment_cons_check}
\end{align}
which is the reflection of momentum conservation within the parton shower functional. This is exact for any system of massless particles. Configuration by configuration, energy and momentum conservation also require that
\begin{align}
    \vev{1,E_m} = Q \, , \quad {\rm and } \quad \vev{\hat n,E_m} = 0 \, .
\end{align}
Using these relations, we can write the parton shower functional as
\begin{align}
    \PSL[\la] = \int \df \sigma_2 ~ e^{-\vev{\la,E_2}} + \int \df \sigma_3 ~ e^{-\vev{\la,E_3}} + \mathcal{O}\left(\as^2\right) \, ,
\end{align}
where $\df \sigma_2$ is the differential Born $\gamma^* \rightarrow q \bar q$ cross section plus the one-loop correction and $\df \sigma_3$ is the tree-level $\gamma^* \rightarrow q \bar q g$ differential cross section. Both terms contain infrared singularities and so we rearrange them to make KLN cancellations manifest,
\begin{align}
    \PSL[\la] = \int ( \df \sigma_2 + \df \sigma_3)~ e^{-\vev{\la,E^B_2}} - \int \df \sigma_3 ~ e^{-\vev{\la,E^B_2}} \left[1 - e^{-\vev{\lambda, \Delta E}}\right] + \mathcal{O}\left(\as^2\right) \, ,
\end{align}
where $\Delta E = E_3 - E^B_2$ and $E^B_2$ is the Born kinematics. Without approximation, both integrals are now manifestly finite. However, we will need to specify a momentum map for reconstructing Born kinematics from the three-parton phase space in order to compute the terms $\df \sigma_3 ~ e^{-\vev{\la,E^B_2}}$. If applied consistently, the choice of momentum map must ultimately cancel from $\PSL[\la]$, only affecting intermediate terms. We will specify the map as and when it becomes necessary.

We will assume that $\la(\nh_i)$ is suitably well behaved around the Born configuration; in particular, $\partial \la(\nh_q^B)$, $\partial^2 \la(\nh_q^B)$, $\partial \la(\nh_{\bar q}^B)$, and $\partial^2 \la(\nh_{\bar q}^B)$ are each well defined and not parametrically large.\footnote{We note that it is not necessary, in general, for $\la(\nh_i)$ to be twice differentiable. This assumption is adopted here only to simplify the presentation of the results. Our analysis can be extended to accommodate discontinuities at $\nh_q^B$ and $\nh_{\bar q}^B$ through the careful introduction of directional derivatives. Such an extension does not significantly modify the final result; rather, it only complicates equations~\eqref{eq:soft_log_scale} and \eqref{eq:col_log_scale}. This would be necessary for observables which depend on $|\sin\theta|$ where $\theta$ is the angle with respect to the Born axis.} We note that, at this order, the requirement that $\la(\nh_i)$ be sufficiently well behaved is, in practice, very similar to requiring that it describe a recursively infrared- and collinear-safe observable in Laplace space \cite{Banfi:2004yd}.

\subsection{The real emission term}

Here we focus on computing 
\begin{align}
    R[\la] = \int \df \sigma_3 ~ e^{-\vev{\la,E^B_2}} \left[1 - e^{-\vev{\lambda, \Delta E}}\right] \, ,
\end{align}
at tree level in the soft and collinear limits to expose the logarithmic terms \cite{Dokshitzer:1991wu,Ellis:1996mzs}. We start by writing
\begin{align}
    \df \sigma_3 = \df \sigma_B \sum_{l = q , \bar q} \frac{\as(\mu)}{2\pi} \frac{\df \cos \theta_l}{1 - \cos \theta_l} \frac{\df \phi_l}{2\pi} ~ \df z_l ~ P_{gq}(z_l) \, , \quad {\rm and} \quad P_{gq}(z) = C_{\rm F} \frac{1 + (1-z)^2}{z} \ ,
\end{align}
where $\theta_l$ is the angle between the emitted gluon and its parent leg indexed by $l$, and $\phi_l$ the azimuth around that leg. $z_l$ is the energy fraction of the emitted gluon so that $\omega_g = z_l Q/2$. We choose to work with the energy fraction rather than the light-cone fraction since the weight from the measurement is defined in terms of energies, and since it has a simple phase-space boundary in the soft and collinear limits $z_l \in [0,1]$. $\df \sigma_B$ is the 2-parton differential Born cross section.

To avoid double counting, we partition $\df \sigma_3 $ into a soft term and a hard collinear term:
\begin{align}
    \df \sigma^{\rm soft}_3 = \df \sigma_B \sum_{l = q , \bar q} \frac{C_{\rm F} \, \as(\mu)}{\pi} \frac{\df \cos \theta_l}{1 - \cos \theta_l} \frac{\df \phi_l}{2\pi} ~ \frac{\df z_l}{z_l}  \, ,
\end{align}
and
\begin{align}
    \df \sigma^{\rm h.c.}_3 =  - \df \sigma_B \sum_{l = q , \bar q} \frac{C_{\rm F} \, \as(\mu)}{2 \pi} \frac{\df \cos \theta_l}{1 - \cos \theta_l} \frac{\df \phi_l}{2\pi} ~ (2 - z_l) \, \df z_l  \, .
\end{align}

\paragraph{The soft term.} Our starting point is
\begin{align}
    &R^{\rm soft}[\la] = \int \df \sigma^{\rm soft}_3 ~ e^{-\vev{\la,E^B_2}} \left[1 - e^{-\vev{\lambda, \Delta E}}\right] \nonumber \\
    & = \frac{C_{\rm F} \, \as(\mu)}{\pi}  \int \df \sigma_B ~ e^{-\vev{\la,E^{B}_2}}  ~  \sum_{l = q , \bar q} \int_{-1}^{1} \frac{\df \cos \theta_l}{1 - \cos \theta_l} ~ \int \frac{\df \phi_l}{2\pi}~ \int_0^1\frac{\df z}{z} \left[1 - e^{-\vev{\lambda, \Delta E}}\right] \, ,
\end{align}
where we have commuted $\vev{\la,E^B_2}$ past the gluon integrals so that it weights just the Born cross section. We will work with the quark as the emitter ($l = q$) term, then use symmetry.

Now we must compute $\vev{\lambda, \Delta E}$. This step requires more care than the usual $z \rightarrow 0$ limit since naively replacing $\vev{\lambda, \Delta E} = \lambda(\hat n_g) z Q/2$ leaves the $\theta_q = 0, \pi$ singularities unregulated. Indeed, it must be the case that $\vev{\lambda, \Delta E} \rightarrow 0$ as $\theta_q \rightarrow 0 , \pi$ to satisfy both momentum conservation and collinear safety. Therefore, consider the usual momentum conservation with 3-body massless kinematics:
\begin{align}
    \omega_q + \omega_{\bar q} + \omega_g = Q \, , \quad {\rm and} \quad (p_q + p_{\bar q} + p_g)^2 = Q^2 \, ,
\end{align}
from which
\begin{align}
    2 p_i \cdot p_j = Q^2 - 2 p_k \cdot (p_i + p_j) = Q^2 - 2 p_k \cdot (q - p_k) = Q^2 - 2 \omega_k Q \, , \label{eq:exact_Kin}
\end{align}
for $(i \neq j \neq k) \in \{q, \bar q, g\}$. As a result,
\begin{align}
    \vev{\lambda, E_3} = \lambda(\hat n_g) \frac{zQ}{2} + \lambda(\hat n_q) \frac{Q^2 - zQ \omega_{\bar q}(1 + \cos \theta_q )}{2 Q}  + \lambda(\hat n_{\bar q}) \frac{Q^2 - zQ \omega_{q} (1 - \cos \theta_q )}{2 Q} \, .
\end{align}
We can now take the soft limit. For the soft momentum map, we specify that $\omega_{q , \bar q} = Q/2$ wherever corrections to these values enter at $\cO(z^2)$ in $\vev{\lambda, E_3}$, and $\hat n_{\bar q} = \hat n_{\bar q}^B = - \hat n_{q}^B$, whilst $\hat n_{q}$ is deflected to conserve transverse momentum,
$$
\hat n_q = \frac{\hat n_q^B - z \hat n_g}{\sqrt{1 + z^2 - 2 z \cos \theta_q}} \, .
$$ This momentum map choice will ultimately cancel provided we handle the collinear terms consistently. We therefore find that in the soft limit
\begin{align}
    &\vev{\lambda, \Delta E} = \vev{\lambda, E_3} - \vev{\lambda, E_2^B} \nonumber \\
    &= \frac{zQ}{2}\left[\lambda(\hat n_g) - \lambda(\hat n^B_q) \frac{1 + \cos \theta_q}{2}  - \lambda(\hat n_{\bar q}^B) \frac{1 - \cos \theta_q}{2} - \sin \theta_q ~ \partial_{\bot} \lambda(\hat n^B_q) \right] + \Ord{z^2} \, ,
\end{align}
where $\partial_{\bot}$ is the derivative in the celestial sphere at $\hat n^B_q$ towards $\hat n_g$. These derivatives are defined carefully when we discuss the collinear piece, where they play a more significant role. This result shows why the additional care was needed. The angular terms which ensure collinear
safety are $\mathcal{O}(z)$, as is the naive contribution. 

Finally, the $z$ integral can be performed using the following standard integral which defines the \emph{entire exponential integral},
\begin{align}
    {\rm Ein}(a) = \int_0^1 \frac{\df t}{t} \left( 1- e^{-at} \right) = \gamma_E + \ln a + \int_a^{\infty} \df s ~  \frac{e^{-s}}{s} \, ,
    \label{eq:comp_exp_int}
\end{align}
whose two limiting behaviours are given by,
\begin{align}
    {\rm Ein}(a) = \gamma_E + \ln a + \mathcal{O} \! \left(\frac{e^{-a}}{a}\right) ~~ {\rm for} ~ a \gg 1 \, ,
    \qquad
    {\rm Ein}(a) = a + \mathcal{O}(a^2) ~~ {\rm for} ~ a \ll 1 \, .
    \label{eq:Ein_limits}
\end{align}
The small-argument behaviour is what makes ${\rm Ein}$ vanish wherever the measurement becomes blind,
and so is what regulates the angular integrals over the collinear regions. We find that
\begin{align}
    &R^{\rm soft}[\la] =  \int \df \sigma_B ~ e^{-\vev{\la,E^{B}_2}}  ~ \int \frac{\df^2 \hat n_g}{4 \pi} \, \sum_{l=q, \bar q} \mathcal{W}(\hat n_g, \hat n^B_l) ~ {\rm Ein} \! \left(\frac{Q}{2 \bar \mu(\hat n_g, \hat n_l^B)}\right) \, ,
\end{align}
where 
\begin{align}
    \mathcal{W}(\hat n_g, \hat n^B_l) = \frac{2 C_{\rm F} \, \as(\mu)}{\pi}\frac{1}{1-\cos \theta_l} \, ,
\end{align}
with $\theta_l$ measured from the parent leg, and 
\begin{align}
    \bar \mu(\hat n_g, \hat n_q^B)^{-1} = \lambda(\hat n_g) - \lambda(\hat n_q^B) \frac{1 + \cos \theta_q}{2}  - \lambda(- \hat n_{q}^B) \frac{1 - \cos \theta_q}{2} - \sin \theta_q ~ \partial_{\bot} \lambda(\hat n^B_q)  \, . \label{eq:soft_log_scale}
\end{align}

\paragraph{The hard collinear term.} This term is evaluated in the collinear limit. We therefore replace
\begin{align}
    \frac{\df \cos \theta_l}{1 - \cos \theta_l} \rightarrow \frac{\df \theta_l^2}{\theta_l^2}
\end{align}
and treat the celestial sphere as locally flat around an emitter $l$ so that the gluon emission axis is specified by a 2D Cartesian vector $\vec{\theta}_l = \theta_l \hat n_{\bot}$ which satisfies $\hat n_{\bot} \cdot \hat n_l = 0$ and $\hat n_{\bot}^2=1$. The hard collinear term can be written as
\begin{align}
    R^{\rm h.c.}[\lambda] = - \frac{C_{\rm F} \, \as(\mu)}{2 \pi}  \int \df \sigma_B ~ e^{-\vev{\la,E^{B}_2}} \sum_{l = q , \bar q} \int_0^1 \frac{\df \theta_l^2}{\theta_l^2} \frac{\df \phi_l}{2\pi} \, \df z_l ~ (2 - z_l)  \left[1 - e^{-\vev{\lambda, \Delta E}}\right] \, .
\end{align} 

Just as in the soft sector, the entire difficulty will be in finding the leading collinear limit of $\vev{\lambda, \Delta E}$. Again, we start from exact kinematics, and fix the parent of the gluon as the quark. From equation~\eqref{eq:exact_Kin}, $\omega_{\bar q} = Q/2 - p_q \cdot p_g / Q$ and so 
\begin{align}
    \vev{\lambda, E_3} = \lambda\left(\hat n_g \right) \omega_g + \lambda\left(\hat n_q \right) \omega_q + \lambda\left( \hat n_{\bar q} \right) \left(\frac{Q}{2} - \frac{p_q \cdot p_g}{Q} \right) \, .
\end{align}
Momentum conservation requires that
\begin{align}
    \vec p_q + \vec p_g = - \, \omega_{\bar q} \, \hat n_{\bar q} = - \left(\frac{Q}{2} - \frac{p_q \cdot p_g}{Q}\right) \hat n_{\bar q} \, ,
\end{align}
whilst energy conservation gives
\begin{align}
    \omega_q + \omega_g = Q - \omega_{\bar q} = \frac{Q}{2} + \frac{p_q \cdot p_g}{Q} \, ,
\end{align}
the two differing by the invariant mass of the collinear pair. It is the energy sum which weights the
measurement, and we use it to define the energy fraction $z$ of the emission, $\omega_g = z(\omega_q+\omega_g)$.
Additionally, let 
$$
\hat n_g = \frac{\alpha \vec{\theta}_q - \hat n_{\bar q}}{\sqrt{1+\alpha^2 \theta^2_q}} \quad {\rm and} \quad \hat n_q = \frac{\beta \vec{\theta}_q - \hat n_{\bar q}}{\sqrt{1+\beta^2 \theta^2_q} } \, ,
$$ so that $\alpha - \beta = 1$ and the splitting has a complete opening angle $\theta_q$. Therefore,
\begin{align}
    &\vev{\lambda, E_3} = \nonumber \\
    &\left(\frac{Q}{2} + \frac{p_q \cdot p_g}{Q} \right) \left[ \lambda\left(\frac{\alpha \vec{\theta}_q - \hat n_{\bar q}}{\sqrt{ 1+\alpha^2 \theta^2_q }}\right) z  
    + \lambda\left(\frac{\beta \vec{\theta}_q - \hat n_{\bar q}}{ \sqrt{1+\beta^2 \theta^2_q }}\right) (1-z) \right]  + \lambda\left( \hat n_{\bar q} \right) \left(\frac{Q}{2} - \frac{p_q \cdot p_g}{Q} \right) \, .
\end{align}
Let $\partial_{\bot}$ be the derivative in the celestial sphere at the point $\hat n$ in the direction of $\vec \theta_q$,
\begin{align}
    &\partial_{\bot} = \hat n_{\bot}^i (\delta_i^j - \hat n_i \hat n^j) \frac{\partial}{\partial \hat n^j} = \hat n_{\bot}^i   \frac{\partial}{\partial \hat n^i} \, , \nonumber \\ 
    &\partial_{\bot}^2 = \hat n_{\bot}^i  \hat n_{\bot}^j  \left[  \frac{\partial}{\partial \hat n^i} \frac{\partial}{\partial \hat n^j} - \delta_{ij} ~ \hat n^k \frac{\partial}{\partial \hat n^k}\right] \, ,
\end{align} 
and expand in $\theta_q$,\footnote{These derivatives can be naturally expressed in spherical polar coordinates where they correspond to derivatives for fixed radius.} noting that the denominators do not contribute as they only rescale the unit vectors but do not change direction,
\begin{align}
    \vev{\lambda, E_3} = 
    &\frac{Q}{2}\Bigg[ 1 + z\left(\alpha \theta_q \partial_{\bot} + \frac{\alpha^2 \theta_q^2}{2} \partial^2_{\bot}\right)  + (1-z) \left(\beta \theta_q \partial_{\bot} + \frac{\beta^2 \theta_q^2}{2} \partial^2_{\bot}\right) \Bigg] \lambda\left(- \hat n_{\bar q}\right) \nonumber\\
    &+ \lambda\left(\hat n_{\bar q} \right) \left(\frac{Q}{2} - \frac{z(1-z)\theta_q^2 Q}{8} \right) + \lambda\left( - \hat n_{\bar q} \right) \frac{z(1-z)\theta_q^2 Q}{8} + \mathcal{O}\left(\theta_q^3, \theta_q^3 \partial^3_{\bot}\lambda\left(- \hat n_{\bar q}\right)\right) \, .
\end{align}
Momentum conservation transverse to the $\hat n_{\bar q}$ axis requires that $z \alpha + (1-z)\beta = 0$ and so the linear terms cancel. We subtract from this the Born term to find
\begin{align}
    \vev{\lambda, \Delta E} = 
    &\frac{Q}{4}z(1-z) \theta_q^2 \partial^2_{\bot} \lambda\left(- \hat n_{\bar q}\right) + \frac{Q}{2} \left(\lambda(\hat n_{\bar q})- \lambda(\hat n^B_{\bar q}) + \lambda( - \hat n_{\bar q})- \lambda( - \hat n^B_{\bar q})\right)\nonumber\\
    &+ \left(\frac{z(1-z)\theta_q^2 Q}{8} \right) \left(\lambda\left( - \hat n_{\bar q} \right) - \lambda\left(\hat n_{\bar q} \right) \right) + \mathcal{O}\left(\theta_q^3, \theta_q^3 \partial^3_{\bot}\lambda\left(- \hat n_{\bar q}\right)\right) \, .
\end{align}
We can now fix our collinear momentum map. Cancellations clearly occur if we choose to align the Born $\bar q$ axis with $\hat n_{\bar q}$, giving,
\begin{align}
    \vev{\lambda, \Delta E} = \frac{Q}{4}z(1-z) ~ \theta_q^2 ~ \cK(\lambda; \hat n^B_{q}, \phi_q) + \mathcal{O}(\theta_q^3) \, ,
\end{align}
where
\begin{align}
    \cK(\lambda; \hat n^B_{q},\phi_q) = \partial^2_{\bot} \lambda\left(\hat n^B_{q}\right) + \frac{1}{2} \left(\lambda\left(\hat n^B_{q} \right) - \lambda\left( - \hat n^B_{q} \right) \right)\, , \label{eq:col_log_scale}
\end{align}
which depends only on the Born kinematics and the azimuth of the collinear emission (via the transverse derivatives). However, the choice of momentum map only cancels from $\PSL[\lambda]$ if applied consistently across the calculation, and we must therefore check for consistency against the soft piece. We must therefore check that the soft map reduces to this collinear map in the limit where they intersect. Expanding the soft weight in the small $\theta_q$ limit, we find
\begin{align}
    &\frac{z Q}{2} \bar \mu(\hat n_g, \hat n_q^B)^{-1} \nonumber \\
    &= \frac{z Q}{2} \left[ \lambda\left( \frac{\hat n_q^B + \vec\theta_q}{\sqrt{1 + \theta_q^2}} \right) - \lambda(\hat n_q^B) \left(1 - \frac{\theta_q^2}{4} \right) - \lambda(- \hat n_{q}^B) \frac{\theta_q^2}{4}  -  \theta_q \partial_{\bot} \lambda(\hat n^B_q)\right] + \mathcal{O}(\theta_q^3, z^2)\, , \nonumber \\ 
    & = \frac{Q}{4}z ~ \theta_q^2 ~ \cK(\lambda; \hat n^B_{q}, \phi_q) + \mathcal{O}(\theta_q^3, z^2)\, ,
\end{align}
and so the two are consistent.

The angular integral can now be performed using equation~\eqref{eq:comp_exp_int},
\begin{align}
    &R^{\rm h.c.}[\lambda] = \nonumber \\ 
    &~~~ - \frac{\as(\mu) C_{\rm F}}{2 \pi}  \int \df \sigma_B ~ e^{-\vev{\la,E^{B}_2}} \sum_{l = q , \bar q} \int \frac{\df \phi_l}{2\pi} \int_0^1 \df z ~ (2-z) ~ {\rm Ein} \left(\frac{z(1-z)Q}{4} \cK(\lambda; \hat n^B_{l},\phi_l) \right) \, ,
\end{align}
where we have dropped the explicit error terms. The hard collinear logarithm is therefore generated by a sum of two pieces, both within $\cK$: a curvature term sensitive to the curvature of the source near the Born configuration, and a forwards-backwards asymmetry of the source.

\paragraph{Overlap terms.} So far we have computed the soft piece in the soft limit and the hard collinear piece in the collinear limit. However, the soft piece also includes a collinear pole. We should therefore check that we have not missed collinear physics within the soft term by only keeping $\cO(z)$ dependence on the kinematics over the complete angular range. This means computing the soft piece in the collinear limit and subtracting the soft piece in the simultaneous soft and collinear limits. Performing this calculation returns a term proportional to
\begin{align}
	\int_0^1\frac{\df z}{z}\left[\mathrm{Ein}\left(\frac{z(1-z)Q}{4}\cK(\lambda;\hat n_l^B,\phi_l )\right)-\mathrm{Ein}\left(\frac{z Q}{4}\cK(\lambda;\hat n_l^B,\phi_l ) \right)\right] \, 
\end{align}
after the $\theta_l$ integration. In the limit that $Q\cK \gg 1$ this converges to a constant and so it is not logarithmically enhanced.\footnote{Moreover, the constant depends on how we define the soft and collinear kinematics and is entirely beyond the control of our log accurate calculation.} Consequently, $R^{\rm h.c.}$ and $R^{\rm soft}$ contain the complete logarithmic structure.

\subsection{The complete result}

Collating what we have so far,
\begin{align}
    \PSL[\la] = \int ( \df \sigma_2 + \df \sigma_3)~ e^{-\vev{\la,E^B_2}} - R^{\rm soft}[\lambda] - R^{\rm h.c.}[\lambda] \, .
\end{align}
What remains is to compute the virtual contribution and to expose the logarithms. The $\vev{\la,E^B_2}$ weight in the first term is trivial with respect to integration over $\df \Pi_g$. Consequently, the inclusive cross section is exposed,
\begin{align}
    \int ( \df \sigma_2 + \df \sigma_3)~ e^{-\vev{\la,E^B_2}} = \int \df \sigma_B ~ e^{-\vev{\la,E^{B}_2}} \left( 1 + \frac{3 \as C_{\rm F}}{4 \pi} \right) + {\rm non\hbox{-}logarithmic} + \mathcal{O}(\as^2) \, ,
\end{align}
where the additional non-logarithmic terms depend only on how we define the momentum map from the three-parton phase space to the Born phase space. These terms cancel for unpolarised production. We artificially keep the non-logarithmic $3 \as C_{\rm F} / 4 \pi$ term so that it is clear that setting $\lambda = 0$ returns the complete cross section. Combining the three terms,
\begin{align}
    &\PSL[\la] =  \int \df \sigma_B ~ e^{-\vev{\la,E^{B}_2}} \Bigg[ \left( 1 + \frac{3 C_{\rm F} \, \as(\mu)}{4 \pi} \right) 
    - \int \frac{\df^2 \hat n_g}{4 \pi} \,\sum_{l=q, \bar q} \mathcal{W}(\hat n_g, \hat n^B_l) ~ {\rm Ein} \! \left(\frac{Q}{2 \bar \mu(\hat n_g, \hat n_l^B)}\right) \nonumber \\
    &+ \frac{\as(\mu) C_{\rm F}}{2 \pi} \sum_{l = q , \bar q} \int \frac{\df \phi_l}{2\pi} \int_0^1 \df z ~ (2-z) ~ {\rm Ein} \left(\frac{z(1-z)Q}{4} \cK(\lambda; \hat n^B_{l},\phi_l) \right)\Bigg] + \mathcal{O}(\as^2, \as L^0)\, . \label{eq:bottomline}
\end{align}
This is the bottom-line result for $\PSL[\la]$ at one loop for twice-differentiable $\la$ with convergent derivative expansions around the Born configuration. One cannot robustly progress further without picking a source $\lambda$. It can be checked that equation~\eqref{eq:bottomline} satisfies equation~\eqref{eq:moment_cons_check}. 

In the ``large-log'' limit, where $\bar \mu \ll Q$ and/or $Q\cK\gg 1$, the large-argument form in equation~\eqref{eq:Ein_limits} applies and our result simplifies to
\begin{align}
    \PSL[\la] = & \int \df \sigma_B ~ e^{-\vev{\la,E^{B}_2}} \Bigg[ \left( 1 + \frac{3 C_{\rm F} \, \as(\mu)}{4 \pi} \right) 
    - \int \frac{\df^2 \hat n_g}{4 \pi} \, \sum_{l=q, \bar q} \mathcal{W}(\hat n_g, \hat n^B_l) ~ \ln \! \left(\frac{e^{\gamma_E} Q}{2 \bar \mu(\hat n_g, \hat n_l^B)}\right) \nonumber \\
    &+ \frac{3\as(\mu) C_{\rm F}}{4 \pi} \sum_{l = q , \bar q} \int \frac{\df \phi_l}{2\pi} ~ \ln \left( e^{\gamma_E} Q \cK(\lambda; \hat n^B_{l},\phi_l) \right)\Bigg] + \mathcal{O}(\as^2, \as L^0)\, . 
\end{align}
This clearly exposes the logarithmic structure at one loop, namely a double logarithm and a set of single logarithms. It is, however, only a heuristic expression and should not be used for direct computations, since collinear safety is ensured by the behaviour ${\rm Ein}(a) = a + \mathcal{O}(a^2)$ for $a \ll 1$.

For any observable whose source $\lambda$ does not satisfy the differentiability and convergence requirements necessary for equation~\eqref{eq:bottomline}, one must return to the more general form:
\begin{align}
    \PSL[\la] = & \int \df \sigma_B ~ e^{-\vev{\la,E^{B}_2}} \nonumber \\
	& \times\Bigg[ \left( 1 + \frac{3 C_{\rm F} \, \as(\mu)}{4 \pi} \right) 
    - \frac{C_{\rm F} \, \as(\mu)}{\pi} \sum_{l = q , \bar q} \int_{-1}^{1} \frac{\df \cos \theta_l}{1 - \cos \theta_l} ~ \int \frac{\df \phi_l}{2\pi}~ \int_0^1\frac{\df z_l}{z_l} \left[1 - e^{-\vev{\lambda, \Delta E}}\right]  \nonumber \\
    &~~~~~~ + \frac{C_{\rm F} \, \as(\mu)}{2 \pi} \sum_{l = q , \bar q} \int_0^1 \frac{\df \theta_l^2}{\theta_l^2} \frac{\df \phi_l}{2\pi} \, \df z_l ~ (2 - z_l)  \left[1 - e^{-\vev{\lambda, \Delta E}}\right] \Bigg]  + \mathcal{O}(\as^2, \as L^0)\, .
\end{align}
One such observable is the back-to-back limit of the energy--energy correlator \cite{Dokshitzer:1999sh,Moult:2018jzp,Ebert:2020sfi}, whose logarithmic asymptotics result in sensitivity to $\partial^n \lambda \sim (\pi - \theta)^{-n}$ where $\theta$ is the correlator opening angle and its deflection from $\pi$ is induced by soft recoil.

\subsection{Example: thrust}
\label{sec:thrust}

To illustrate the application of equation~\eqref{eq:bottomline}, we will now use it to compute thrust. Thrust is given by
\begin{align}
	\tau = 1-T = 1 - \max_{\hat n_\tau} \frac{\sum_i | \hat n_\tau \cdot \vec{p}_i |}{\sum_i |\vec{p}_i|} = \frac{\sum_i \omega_i - \max_{\hat n_\tau} \sum_i | \hat n_\tau \cdot \hat n_i | ~ \omega_i}{Q} \, .
\end{align}
In Laplace space, thrust weights the measurement by
\begin{align}
	\exp\left( - \nu \frac{\sum_i \omega_i - \max_{\hat n_\tau} \sum_i | \hat  n_\tau \cdot \hat n_i | ~ \omega_i}{Q} \right) \, ,
\end{align}
where $\nu$ is the Laplace conjugate variable to the thrust parameter $\tau = 1-T$. With $\as L$ accuracy, we can align the thrust axis with the Born $q \bar q$ axis allowing us to drop the energy-dependent maximisation. Therefore, up to non-logarithmic terms, the one-loop thrust Laplace-space distribution is computed from the source
\begin{align}
    \lambda_{\tau}(\hat n ) = \nu (1 - |\hat n \cdot \hat n_\tau|)/Q
\end{align}
with $ \hat n_\tau$ the thrust axis, which we take to be the Born $q \bar q$ axis.\footnote{We note that dropping the maximisation is not a general feature of the thrust axis at $\as L$ accuracy. Rather, it is a feature of the thrust observable itself and its quadratic dependence on axis deflections. The maximisation must be treated more carefully for observables that depend linearly on deflections from the thrust axis, such as the broadening defined relative to the thrust axis.}

Throughout this subsection we write
\begin{align}
    L \; \equiv \; \ln \nu + \gamma_E \, ,
\end{align}
which is the combination in which the logarithms of a Laplace transform naturally appear.

Looking first at the soft term, $\lambda_{\tau}(\hat n^B_q ) = \lambda_{\tau}(\hat n^B_{\bar q} ) = \partial_{\bot} \lambda_{\tau}(\hat n^B_q ) = \partial_{\bot} \lambda_{\tau}(\hat n^B_{\bar q} ) = 0$ and so $\bar \mu(\hat n_g, \hat n_l^B) = \mu(\hat n_g)$. Hence,
\begin{align}
    R^{\rm soft}[\lambda_\tau] & = \sigma_B \int \frac{\df^2 \hat n_g}{4 \pi} \, \left(\mathcal{W}(\hat n_g, \hat n^B_q) + \mathcal{W}(\hat n_g, \hat n^B_{\bar q}) \right)~ {\rm Ein} \! \left(\frac{\nu}{2} (1 - |\hat n_g \cdot \hat n_\tau|)\right) \, ,\nonumber \\
    &= \frac{\as(\mu) C_{\rm F} \sigma_B }{ \pi} L^2 + \mathcal{O}(\as L^0) \, ,
\end{align}
up to terms suppressed by powers of $1/\nu$. Had $\gamma_E$ been discarded, the single-logarithmic
term $2 \gamma_E \ln \nu$ contained in $L^2$ would have been lost.
Turning to the hard collinear piece, in this limit $\lambda_{\tau} = \nu \vec{\theta}^{\, 2}/(2Q)$ where $\vec{\theta}$ is the angle to the thrust axis. There is no forwards-backwards asymmetry and so the only contributing term is the double derivative which gives $ \partial_{\bot}^2 \lambda_{\tau}(\hat n_\tau) = \nu / Q$. Therefore, the hard collinear piece is
\begin{align}
    R^{\rm h.c.}[\lambda_\tau] = - \frac{3\as(\mu) C_{\rm F} \sigma_B }{2 \pi} L  + \mathcal{O}(\as L^0)\, ,
\end{align}
where here, unlike in the soft term, the distinction between $L$ and $\ln \nu$ is an
$\mathcal{O}(\as L^0)$ effect. Combining these terms provides the known one-loop computation of logarithmic
terms from the $e^+e^-$ thrust distribution in Laplace space \cite{Catani:1992ua}:
\begin{align}
	\int \df \tau ~ \frac{\df \sigma }{\df \tau} e^{-\nu \tau} = \PSL[\lambda_\tau] = \sigma_B - \frac{\as(\mu) C_{\rm F} \sigma_B }{ \pi} L^2 +  \frac{3\as(\mu) C_{\rm F} \sigma_B }{2 \pi} L  + \mathcal{O}(\as L^0, \as^2) . 
\end{align}

\section{A toy model factorisation for $\PSL$}
\label{sec:toy}

Having computed $\PSL$ at fixed order, we will now consider its all-orders structure. We will discuss a toy model for the hadronic final state of an $e^+ e^-$ collision with just enough structure for us to highlight some key all-orders features of $\PSL$.

For our toy model, we start by assuming the Hamiltonian for the evolution of the $e^+e^-$ final state contains only two decoupled sectors, a hard pseudo-jet sector containing an unspecified number of pseudo-jets and a soft sector resolving only the directions of the hard pseudo-jets. The soft sector is the only low scale degree of freedom, and so in the low scale effective theory
\begin{align}
    \bra{X}\eps_\mu(q)\int\df^4x\;e^{iq\cdot x}\,J^\mu(x)\,\ket{0} \mapsto \sum_m & \bra{X_s;\{b\}} S_{1}^{b_1 a_1}(\nh_1) \cdots S_{m}^{b_m a_m}(\nh_m) \ket{0_s} \nonumber \\
    & \times \cM^{a_1 \dots a_m}_m(\{x\}_m,\{\hat n \}_m) \, ,
\end{align}
where $\cM_m$ is the matching coefficient of the soft degrees of freedom onto a QCD system created from $J$ with $m$ pseudo-jets (playing the role of a hard matrix element). Each pseudo-jet ($i$) is described by its direction, $\hat n_i$, and energy fraction $x_i=2\omega_i/Q$. $S_i$ is a soft Wilson line which organises, in the low-energy theory, the leading eikonal degrees of freedom around a pseudo-jet $i$ \cite{Bauer:2000yr,Bauer:2001yt},
\begin{align}
    S^{b a}_i (\nh_i) = \Pexp\left( i g_{\rm s} \int_0^\infty \df t ~ n_i \cdot A^c_s(t \, n_i) ~ T^c_{R(i)}\right)^{ba} \, ,
\end{align}
where $T_{R(i)}$ is the generator in the representation $R$ appropriate to the pseudo-jet leg $i$ and $n_i = (1, \hat n_i)$ lies in the direction of $i$. $X_s$ is a soft final state and $\ket{0_s}$ the soft vacuum. We will drop the arguments on the Wilson lines for compactness, $S^{b a}_i (\nh_i) \equiv S^{b a}_i$.

Within this effective description, $\PSL$ factorises simply as
\begin{align}
    \PSL[\lambda] = & \sum_m \left(\prod_{i \leq m} \int  \df^2 \hat n_i \int_0^1 \df x_i  \right)  \cH^{{a_1 \cdots a_m} , {a'_1 \cdots a'_m}}_m(\{x\}_m,\{\hat n \}_m)  \nonumber \\ 
    & \times \bra{0_s} ( S_{m}^{b_m a'_m} )^\dagger \cdots ( S_{1}^{b_1 a'_1} )^\dagger e^{-\vev{\lambda, E_m + \cE_s}} S_{1}^{b_1 a_1} \cdots S_{m}^{b_m a_m} \ket{0_s}\, ,
\end{align}
where 
\begin{align}
    \cH^{{a_1 \cdots a_m} , {a'_1 \cdots a'_m}} = (\cM^{a'_1 \dots a'_m}_m)^\dagger \cM^{a_1 \dots a_m}_m \, ,
\end{align}
is the hard density matrix, $E_m$ is the recoiling flux of the $m$ pseudo-jets, and $\cE_s$ is the energy-flow operator restricted to the soft sector. This can be more compactly written as
\begin{align}
    \PSL[\lambda] = & \sum_m \Tr \left[ \boldsymbol{\cH}_m(\{x\}_m,\{\hat n \}_m) \otimes_{\{x, \hat n\}} \boldsymbol{S}_m(\{x\}_m,\{\hat n \}_m;\lambda) \right]\, , \label{eq:toy}
\end{align}
where bold symbols are tensors in colour space, over which $\Tr$ is a trace, and $\otimes_{\{x, \hat n\}}$ implies the integration over the shared $\{x\}_m$ and $\{\hat n \}_m$ dependence. Note that the soft matrix element still depends on $\{x\}_m$ through the recoiling $E_m$.

This factorisation for $\PSL$ almost precisely matches that of Becher et al. \cite{Becher:2015hka,Becher:2016mmh,Becher:2021zkk,Becher:2023mtx}, introduced for the resummation of non-global observables. This is not a coincidence. Consider the foundational object in our derivation of $\PSL$, the cell source $\lambda(\hat n) = \lambda_a \chi_a(\hat n)$. Measuring just the flux in this cell is the Laplace transform of measuring a non-global observable where the cell is required to measure some energy $E_a$ (the conjugate variable to $\lambda_a$) and our measurement is otherwise inclusive. In the case studied by Becher et al., the measurement is performed in a patch far from the jet axis, so it is sensitive only to soft radiation, and our toy model is the complete picture at leading power. One can therefore identify a physical realisation of the toy-model factorisation in Eq.~\eqref{eq:toy}. It corresponds to the factorised cross section obtained when making a large number of non-global measurements, all at a similar scale, across a lattice of patches. Taking the continuum limit of this lattice in Laplace space then reproduces the factorisation structure of Eq.~\eqref{eq:toy}.

Following the work of Becher et al., we can renormalise the factorisation leading to a multiplicity off-diagonal evolution for the hard function. 
\begin{align}
    \PSL[\lambda] =  \sum_{m' \geq m} \Tr \bigg[ & \boldsymbol{\cH}_m\left(\{x\}_m,\{\hat n \}_m ; \frac{\mu}{Q}\right) \otimes_{\{x, \hat n\}} \boldsymbol{U}_{m m'}(\{x\}_m,\{\hat n \}_m,\{x'\}_{m'},\{\hat n' \}_{m'}; \mu, \mu_{\rm IR} )  \nonumber \\
    &\otimes_{\{x', \hat n'\}} \boldsymbol{S}_{m'}\left(\{x '\}_{m'},\{\hat n ' \}_{m'};\ln(\lambda \mu_{\rm IR}) \right) \bigg]\, , 
\end{align}
where $\boldsymbol{U}$ is defined in terms of the anomalous dimension of the hard density matrix and mixes pseudo-jet multiplicities
\begin{align}
    \boldsymbol{U}(\underline{\{x\}},\underline{\{\hat n \}}; \mu, \mu_{\rm IR} ) = \Pexp\left( \int_{\mu_{\rm IR}}^{\mu} \frac{\df \mu'}{\mu'} \boldsymbol{\Gamma}^{\rm H} (\underline{\{x\}},\underline{\{\hat n \}}) \right) \, 
\end{align}
where
\begin{align}
    \boldsymbol{\Gamma}^{\rm H} (\underline{\{x\}},\underline{\{\hat n \}})_{mm'} = \boldsymbol{\Gamma}^{\rm H} (\{x\}_m,\{\hat n \}_m, \{x\}_{m'},\{\hat n \}_{m'} )_{mm'} = \boldsymbol{V}_m \delta_m^{m'} + \boldsymbol{R}_m \delta_{m+1}^{m'} + \dots \, .
\end{align}
However, for any non-trivial source $\lambda$, there is no single value of $\mu_{\rm IR}$ which achieves a complete resummation of large logs: if $\lambda(\hat n_k) \gg \lambda(\hat n_l)$ then $\ln [\lambda(\hat n_k)  \mu_{\rm IR}] \gg 1$ when $\ln [\lambda(\hat n_l)  \mu_{\rm IR}] \sim 0$. The renormalisation-group evolution of the hard function is not sufficient to fully constrain all possible large logarithms since $\boldsymbol{S}_{m}$ is not a single-scale matrix element; rather, it depends on a continuum of scales. However, $\boldsymbol{S}_{m}$ also cannot be refactorised within a traditional effective theory, since the continuum of scales necessarily contains a continuum of scale hierarchies which does not permit a power expansion. 

The generalised resummation of large logarithms in $\PSL$ is what a parton shower must accurately achieve and, as we have argued, it goes beyond the most straightforward application of renormalisation-group-improved perturbation theory. Although the discussion in this section uses a simplified toy model, a complete parton shower must reproduce the logarithmic structure it captures in order to resum non-global logarithms. The toy model demonstrates that a successful parton shower must evolve the system more differentially than is permitted by a strict renormalisation-group evolution of the hard functions alone. This additional differential structure is required in order to resolve low-scale radiation, even when one begins with a complete hierarchy of hard functions at different multiplicities.\footnote{We wish to state carefully that we do not find any conflict with the formulations of \cite{Nagy:2017ggp,Nagy:2020rmk,Nagy:2020dvz,Platzer:2022jny}. These formulations treat parton showers as a renormalisation-group evolution in which the renormalisation not only removes singularities, but also removes differential kernels for real emissions that may become singular depending on a measurement boundary. In this way, the renormalisation-group evolution may resum logarithms associated with $\lambda(\hat n)$. Our construction highlights only that the renormalisation group of \cite{Nagy:2017ggp,Nagy:2020rmk,Nagy:2020dvz,Platzer:2022jny} is not, a priori, the same construction as one would obtain in an effective field theory or through minimal subtraction.} We leave the formulation of the generalised logarithmic resummation for $\PSL$, and its implications for parton showering, to future work.

\section{Conclusions}
\label{sec:conclusions}

In this work, we have formulated what an event generator should compute in terms of what a collider detector measures: the distribution of hadronic fluxes. We constructed this distribution from detector operators built from asymptotic hadronic modes, restricted to a commuting family so that their readings possess a joint probability distribution. Events gain the interpretation of direct samples from this distribution. Its functional Laplace transform is the event generator source functional, $\EGL$, and its marginal onto the energy flux, the only infrared- and collinear-safe detector in the family, is the parton shower source functional, $\PSL$. Both are exact, hadron-level and non-perturbative, and unphysical information inaccessible to a detector is integrated out from the outset. KLN cancellations are directly facilitated, and so $\PSL$ is finite with an identifiable logarithmic structure. Moreover, $\PSL$ contains every possible calorimetric measurement, and we demonstrate that event shapes and energy correlators are obtained as particular choices of the source.

We computed the logarithmic structure of $\PSL$ at one loop in $\ee$ annihilation for a general twice-differentiable source. The logarithms are controlled by the source together with its angular derivatives at the Born directions, so that $\PSL$ is directly sensitive to kinematic recoil. We then examined the all-orders structure of $\PSL$ in a toy-model factorisation, which closely matches the factorisation used for non-global observables: $\PSL$ can be viewed as a continuum of non-global measurements. Since the source varies across the celestial sphere, the low-scale matrix element depends on a continuum of scales, and resumming all of its logarithms requires an evolution more differential than a minimal-subtraction-type single-scale renormalisation-group evolution of the hard functions. A parton shower necessarily resolves low-scale radiation, and so can in principle resolve and resum the continuum of scales. Our construction therefore suggests that a parton-shower algorithm goes beyond a simple renormalisation-group evolution generated from a hard function as determined by an effective field theory.

This work opens several directions for future study. Two are of particular significance: to formulate the generalised resummation of the logarithms in $\PSL$ and to realise it with a parton-shower algorithm; and to find a factorisation for $\PSL$ and $\EGL$, thereby facilitating their matching to non-perturbative physics. Such a factorisation would give an analytically controlled formulation of event generation, in which parton showering emerges from the evolution of the factorised components and hadronisation is encoded in low-scale hadronic matrix elements. The operator-based formulation of the low-scale matrix elements may then motivate more sophisticated modelling of event-generator hadronisation.

Finally, the detector-based perspective suggests a possible extension beyond QCD. Energy-flow observables can also be defined in theories such as $\cN=4$ super-Yang–Mills theory, where they may be studied using non-perturbative methods. While the construction of $\EGL$ relies on asymptotic hadronic out states, the energy-flow operator underlying $\PSL$ can alternatively be described as a light transform of the energy–momentum tensor. It can therefore be defined directly through detector correlation functions without relying on a hadronic out-state decomposition or an S-matrix. This could offer new handles on parton showering and event generation by drawing on non-perturbative results from theories adjacent to QCD.

\acknowledgments

I would like to thank Jeff Forshaw, Massimiliano Procura, and Simon Pl\"atzer for discussions which helped inspire this project. Additional thanks also to Jeff Forshaw for comments on the manuscript. Particular thanks also go to Kyle Lee, who has helped to inform my perspective on detector operators.
JH is supported by the Leverhulme Trust as an Early Career Fellow. 
This research was supported in part by grant NSF PHY-2309135 to the Kavli Institute for Theoretical Physics (KITP). 
We thank the Erwin-Schrödinger International Institute for Mathematics and Physics at the University of Vienna for partial support during the Programme ``New Paradigms for Harnessing Quantum Field Theory at Colliders'', July 27 - August 28, 2026.

\appendix

\section{The continuum limit of the calorimeter measure for an atomic flux}
\label{app:atomic}

The discussion we provided of the continuum limit of equation~\eqref{eq:dEm} in section~\ref{sec:PS} relied on equation~\eqref{eq:smoothstep}, which assumes that the energy flux is a smooth field on the celestial sphere. In QCD, the flux realised on every hadronic state is instead the atomic distribution. In this appendix we present $\cD\PS$ explicitly as a measure on atomic flux configurations, and show that it satisfies equation~\eqref{eq:PSdef}.

Assuming a mass gap, and that configurations in which two hadrons arrive from exactly the same direction carry zero weight, the lattice measure in equation~\eqref{eq:dEm} tends, in the continuum limit, to
\begin{align}
	\cD\PS[E]\;&=\;\sum_{n} \int\prod_{i=1}^{n}\df^2\nh_i\;\df E(\nh_i)\;\;
	\bra{\psi}:\!\prod_{i=1}^{n}\de\big(\cE(\nh_i)-E(\nh_i)\big)\!:\sumint_{X_n}\ket{X_n}\vev{X_n|\psi}\,,
	\label{eq:app_PS}\\
	E(\nh)\;&=\;\sum_{i=1}^{n}E(\nh_i)\,\de^{(2)}(\nh_i,\nh)\,,
	\label{eq:app_profile}
\end{align}
where $X_n$ denotes an out state of multiplicity $n$. For each $n$ the term of equation~\eqref{eq:dEm} with $n$ lit cells converges to the $n$-hadron term of equation~\eqref{eq:app_PS} when both are integrated against any continuous function of $\vev{\la_1,E},\dots,\vev{\la_M,E}$ with continuous sources. Here $E(\nh_i)$ is the energy carried by the hadron at $\nh_i$, that is, the coefficient of $\de^{(2)}(\nh_i,\nh)$ in the profile, not the value of the distribution $E(\nh)$ at $\nh_i$. It is necessary to ensure that each hadron is measured only once (i.e. we remove the coincident regions from the product over $\cE$), which we have indicated with the normal ordering symbol:
\begin{align}
	:\!\prod_{i=1}^{n}\de\big(\cE(\nh_i)-E(\nh_i)\big)\!:\,\ket{X_n}\;=\;\prod_{i=1}^{n}\de^{(2)}(\nh_i,\nh_{r_i})\;\de\big(E(\nh_i)-\omega_{r_i}\big)\ket{X_n}\,,
	\label{eq:app_normal}
\end{align}
where the hadrons are ordered, for example by energy, $E(\nh_1)>\dots>E(\nh_n)$, the integrals in equation~\eqref{eq:app_PS} run over this ordering, and $r_i$ is the $i$-th hadron of $X_n$ in the same ordering.
Equation~\eqref{eq:app_PS} reproduces equation~\eqref{eq:PSdef}. Explicitly,
\begin{align}
	&\int\!\cD\PS[E]\;\Phi\big(\vev{\la_1,E},\dots,\vev{\la_M,E}\big)\nn\\
	&\quad=\;\sum_n\sumint_{X_n}\big|\vev{X_n|\psi}\big|^2\int\prod_{i=1}^{n}\Big[\df^2\nh_i\,\df E(\nh_i)\;\de^{(2)}(\nh_i,\nh_{r_i})\;\de\big(E(\nh_i)-\omega_{r_i}\big)\Big]\nn\\
	&\hspace{6em}\times\Phi\Big(\sum_{i}E(\nh_i)\,\la_1(\nh_i),\dots,\sum_{i}E(\nh_i)\,\la_M(\nh_i)\Big)\nn\\
	&\quad=\;\sum_n \sumint_{X_n}\big|\vev{X_n|\psi}\big|^2   ~ \Phi\Big(\sum_{r \in X_n} \omega_{r}\,\la_1(\nh_{r}),\dots,\sum_{r \in X_n}\omega_{r}\,\la_M(\nh_{r})\Big)\nn\\
	&\quad=\;\sum_n\sumint_{X_n}\big|\vev{X_n|\psi}\big|^2 ~ \Phi\big(\vev{\la_1,E_{X_n}},\dots,\vev{\la_M,E_{X_n}}\big)\nn\\
	&\quad=\;\bra{\psi}\,\Phi\big(\vev{\la_1,\cE},\dots,\vev{\la_M,\cE}\big)\,\ket{\psi}\, ,
	\label{eq:app_verify}
\end{align}
where we have used $\Phi\big(\vev{\la_1,\cE},\dots\big)\ket{X}=\Phi\big(\vev{\la_1,E_X},\dots\big)\ket{X}$, and that the sums over multiplicities and states assemble the completeness relation in equation~\eqref{eq:completeness}.

\bibliography{refs}

\end{document}